\documentclass[12pt]{article}

\usepackage{amsmath,amssymb,amsbsy,amsfonts,latexsym,graphicx}
\usepackage{color,array}
\usepackage[nospace,sort,compress]{cite}
\usepackage{emptypage} 

\usepackage[usenames,dvipsnames]{xcolor}
\usepackage[unicode,psdextra]{hyperref}
\hypersetup{
    colorlinks=true, 
    citecolor=ForestGreen,
    linkcolor=RoyalBlue,
    urlcolor=RoyalBlue,
    pdfstartview=FitV,
    breaklinks=true,
    linktocpage=true}
\usepackage{bookmark}

\usepackage{mathtools}

\numberwithin{equation}{section}

\allowdisplaybreaks

\evensidemargin \oddsidemargin

\begin{document}


\begin{titlepage}

\renewcommand{\thefootnote}{\fnsymbol{footnote}}


\begin{flushright}
\end{flushright}

\vspace{15mm}
\baselineskip 9mm
\begin{center}
  {\Large \bf Notes on worldvolume supersymmetries for D-branes \\
              on AdS$_5\times$S$^5$ background}
\end{center}

\baselineskip 6mm
\vspace{10mm}
\begin{center}
Jaemo Park$^a$\footnote{\tt jaemo@postech.ac.kr} and
Hyeonjoon Shin$^{a,b}$\footnote{\tt shin67@postech.ac.kr}
\\[10mm]
  $^a${\sl Department of Physics, POSTECH,\\
        Pohang, Gyeongbuk 37673, South Korea}
\\[3mm]
  $^b${\sl Asia Pacific Center for Theoretical Physics, \\
       Pohang, Gyeongbuk 37673, South Korea}
\end{center}

\thispagestyle{empty}

\vfill
\begin{center}
{\bf Abstract}
\end{center}
\noindent
We revisit the 1/2-BPS D-branes on the AdS$_5\times$S$^5$ background.
Based only on the classification of 1/2-BPS D-branes obtained by
the covariant open string description, we consider various purely
static configurations of D-branes without any worldvolume
flux on the AdS$_5\times$S$^5$ background. Under the covariant $\kappa$
symmetry fixing condition, we investigate which part the spacetime
supersymmetries is preserved on the D-brane worldvolume and obtain the
associated worldvolume supersymmetry transformation rules
to leading order in the worldvolume fluctuating fields.  It is shown
that, for purely static configurations without any worldvolume flux,
only the AdS type D-branes,
in which the AdS radial direction is
one of worldvolume coordinates, are 1/2-BPS.
\\ [15mm]

\vspace{5mm}
\end{titlepage}

\baselineskip 6.6mm
\renewcommand{\thefootnote}{\arabic{footnote}}
\setcounter{footnote}{0}

\tableofcontents

\section{Introduction}
\label{intro}

D-brane in a given background containing  AdS spacetime is an
interesting object to explore.  It carries the information about the open
string sector in the background.  In the AdS/CFT
correspondence \cite{Maldacena:1997re,Gubser:1998bc}, a certain
D-brane configuration on the background involving the AdS spacetime is
related to the defect conformal field
theory (dCFT) \cite{DeWolfe:2001pq,Skenderis:2002vf}.  For example,
a certain D5-brane on the AdS$_5 \times$S$^5$ background is dual to
the three dimensional dCFT of the N=4 SYM theory \cite{Skenderis:2002vf}.
In a suitable approximation, D branes on curved spacetime can be
described by Born-Infeld type action.
The Born-Infeld type action of  supersymmetric D-brane  has an
important symmetry, $\kappa$ symmetry.  By a suitable gauge fixing, one
can obtain the supersymmetric worldvolume theory on the Born-Infeld type
action of D-brane.  In Ref.~\cite{Aganagic:1996pe}, by choosing a static
gauge combined with a suitable $\kappa$ gauge fixing, the supersymmetric
worldvolume theory of D-brane in flat space is obtained, with
the explicit supersymmetric transformation worked out.

One can certainly adopt the same strategy to D-branes on the
background involving the AdS spacetime.  In fact, $\kappa$ symmetric
action of D-branes on the AdS$_5\times$S$^5$ background was obtained using
supercoset approach.  One can guess that again by taking the static gauge
with suitable $\kappa$ gauge fixing condition one can obtain the
worldvolume theories of supersymmetric D-branes on the
AdS$_5\times$S$^5$ background.\footnote{This will be one more useful
example of supersymmetric theories on curved background, which can
find the application to the localization of the gauge theory on a
curved background.}
However there is a subtlety in this program. In \cite{Shin:2015uga} , the
author considered the D3-brane whose world volume spans the four
directions in AdS$_5$ other than the radial direction but could not
find the suitable gauge fixing condition leading to a supersymmetric
worldvolume theory on the D-brane.\footnote{The subtlety arises since the D3 brane whose world volume
we consider is parallel to D3 branes, which make the geometry of AdS$_5\times$S$^5$. If we start from intersecting configuration of a D$p$ brane with D3 branes, which are geometrized , such subtlety  does not arise.} Note that the configuration is supersymmetric
since the D3 brane of interest is parallel to the D3 branes, whose near horizon geometry
turns into AdS$_5\times$S$^5$. Indeed in \cite{Cascales:2004qp}, the D3 brane is shown to satisfy
the generalized calibration, hence is supersymmetric.\footnote{For recent
exploration of the generalized calibration in the AdS backgrounds, see
for example Ref.~\cite{deFelice:2017mhm} and references therein.}
Analogous result is worked out at \cite{Pasti:1999sp}, where they consider M2 brane in  AdS$_4\times$S$^7$
where M2 brane is parallel to M2 branes which make AdS geometry. They show that Killing spinor gauge is
incompatible with the static gauge of the M2 brane worldvolume action.
Given these results, one might wonder if there are some restrictions on the
possible supersymmetric worldvolume thoery on the AdS spacetime.
It turns out that this problem is intimately related to the
classification of supersymmetric D-brane embeddings into the AdS spacetime.

In this paper we are looking for the supersymmetric worldvolume theories
of 1/2-BPS D-branes on the AdS$_5\times$S$^5$ background .
Since the worldvolume supersymmetry is of our concern, it is natural
to take the probe brane analysis.  In order to study the worldvolume
theory of 1/2-BPS
D-branes, we start from the data obtained from the covariant
open string description of supersymmetric D-branes developed in
\cite{Lambert:1999id,Bain:2002tq}.  The covariant
description can be applied to any background if the superstring action
on it is given and, as the first non-trivial application, has led to
the classification of 1/2-BPS D-branes in some plane wave
backgrounds \cite{Bain:2002tq,Hyun:2002xe}.\footnote{The IIB plane wave
has a connection with the AdS$_5\times$S$^5$ background through the
Penrose limit \cite{Blau:2002dy}. One may refer a work
\cite{Freedman:2003kb} done in the plane wave background, which may be
related to the present one.}
As for the AdS$_5\times$S$^5$ background, the classification
has been worked out in \cite{Sakaguchi:2003py,ChangYoung:2012gi} and
Table \ref{tablebps} shows its result.  As a consistency check, we note
that the same data listed in Table \ref{tablebps} have been also obtained
in \cite{Hanazawa:2016lvo} in the context of pure
spinor formalism \cite{Berkovits:2000fe}.

\begin{table}
\begin{center}
\begin{tabular}{c|cccccc}
\hline
  & D(-1) &D1 & D3 & D5 & D7 & D9 \\ \hline\hline
($n$,$n'$) & (0,0) &
\begin{tabular}{c} (2,0) \\ (0,2) \end{tabular} &
\begin{tabular}{c} (3,1) \\ (1,3) \end{tabular} &
\begin{tabular}{c} (4,2) \\ (2,4) \end{tabular} &
\begin{tabular}{c} (5,3) \\ (3,5) \end{tabular} &
-- \\
\hline
\end{tabular}
\caption{\label{tablebps} 1/2-BPS D-branes in the AdS$_5\times$S$^5$
background. $n$ ($n'$) represents the number of Neumann directions in
AdS$_5$ (S$^5$).}
\end{center}
\end{table}

The covariant open string description provides us a definite guideline for
further study of supersymmetric D-branes, although it gives no more
information about D-branes other than the classification data.
Starting from the Table \ref{tablebps}, we  consider all
possible types of corresponding D-brane configurations and use
 a suitable static gauge for the D-brane worldvolume
diffeomorphism.  For the fermionic $\kappa$ symmetry of the D-brane
action, the covariant $\kappa$ symmetry gauge is adopted.
For each of the configurations, we identify the worldvolume
supersymmetry realized on the D-brane worldvolume and obtain the
associated worldvolume supersymmetry transformation rules for the
worldvolume fields.

We  restrict ourselves to the purely static Lorentzian D-brane
configurations without any worldvolume flux.
Here the Lorentzian means that the worldvolume time is identified with that
of the background spacetime.  Thus, the configurations for (0,0) and
(0,2) D-branes of Table \ref{tablebps} are not considered.
As one may realize, there are twelve types of Lorentzian
configurations. Six of them correspond to the AdS type D-branes in which
the AdS radial direction is one of worldvolume coordinates, and the
remaining six are of non-AdS type in which the AdS radial direction is
transverse to the D-brane worldvolume.  We will treat the AdS and non-AdS
type branes separately.
It turns out  that, {\it for purely static configurations without any
worldvolume flux, only AdS type branes admit the supersymmetric
worldvolume theories.} Note that the corresponding D-brane configurations are obtained
from the supersymmmetric intersecting D3$\perp$D$p$ brane configurations in flat
spacetime after turning D3 branes
into AdS geometry \cite{Skenderis:2002vf}.
For non-AdS type
branes, the analysis suggests that world volume fluxes should be turned
on or some motions in transverse directions should be considered to
have the supersymmetric world volume theory.  Our work suggests
that only D-branes tabulated at Table \ref{tablebps} admit
the supersymmetric worldvolume theory. In order to confirm it, the
worldvolume theories of non-AdS type branes should be worked out,
which is beyond the scope of this paper.

The organization of this paper is as follows.  In Sec.~\ref{gen}, we
describe the way of realizing the worldvolume supersymmetry for a given
D-brane configuration after reviewing some necessary elements.
In Sec.~\ref{adsbrane}, we investigate the worldvolume supersymmetry
for the AdS type D-branes.  Then the non-AdS branes are considered in
Sec.~\ref{nonadsbrane}. The discussion with our conclusion follows in
Sec.~\ref{concl-disc}.  Finally, appendix~\ref{app} contains our notation
and convention with the expressions of superfields.

\section{Generalities}
\label{gen}

In this section, we briefly review the AdS$_5\times$S$^5$ background with
its associated Killing spinor and the symmetries of D$p$-brane action.
We then describe how to identify the supersymmetry realized on the brane
worldvolume for a given brane configuration.

\subsection{AdS$_5\times$S$^5$ background}
\label{adsbg}

In the Poincar\'{e} patch coordinates, the metric for the AdS$_5\times$S$^5$
geometry is written as
\begin{align}
ds^2 = u^2 \left[ -(dx^0)^2 + (d\vec{x})^2 \right] + \frac{du^2}{u^2}
      + d\Omega_5^2 \,,
\label{geom}
\end{align}
where $(d\vec{x})^2 = (dx^1)^2 + (dx^2)^2 + (dx^3)^2$ and
$d\Omega_5^2$ is the metric of S$^5$ parametrized by five angular coordinates
$\phi^\alpha$ ($\alpha=1,\dots,5$),
\begin{align}
d\Omega^2_5 =
    (d\phi^1)^2
    + \sin^2 \phi^1
      \big[
           (d\phi^2)^2
           + \sin^2 \phi^2
             \big[
                  (d\phi^3)^2
                  + \sin^3 \phi^2
                    \big[
                         (d\phi^4)^2
                         + \sin^4 \phi^2 (d\phi^5)^2
                    \big]
             \big]
     \big] \,,
\end{align}
with ranges of $0 \le \phi^1, \phi^2, \phi^3, \phi^4 \le \pi$ and
$0 \le \phi^5 \le 2\pi$.
Here, we have taken the common radius $R$ of the AdS$_5$ and S$^5$ to be one,
$R=1$.
The ten dimensional coordinates are aligned as
\begin{align}
X^\mu = \{ x^0, x^1, x^2, x^3, u, \phi^1, \dots, \phi^5 \}
\label{10dcoord}
\end{align}
and from the metric (\ref{geom}) the zehnbein is chosen to be
\begin{align}
e^{0,1,2,3} = u dx^{0,1,2,3} \,, \quad e^4 = \frac{du}{u} \,, \quad
e^{a'} = \prod^{a'-5}_{\alpha=1} \sin \phi^\alpha d\phi^{a'-4} \,.
\label{zbein}
\end{align}
In addition to the metric (\ref{geom}), another constituent of the
AdS$_5\times$S$^5$ background is the Ramond-Ramond five form field strength
given by
\begin{align}
F_5 = 4 e^0 \wedge e^1 \wedge e^2 \wedge e^3 \wedge e^4
     +4 e^5 \wedge e^6 \wedge e^7 \wedge e^8 \wedge e^9 \,.
\label{rr5}
\end{align}

The AdS$_5\times$S$^5$ background composed of (\ref{geom}) and (\ref{rr5})
is maximally supersymmetric.  Its
supersymmetry structure is encoded in the spacetime Killing spinor $\eta^I$,
which is the solution of the spacetime Killing spinor equation
$D_\mu \eta^I(X)= 0$ for the AdS$_5\times$S$^5$
background.\footnote{The explicit form of the covariant derivative
$D_\mu$ can be found in Eq.~(\ref{m2dt}).}
The Killing spinor equation has been solved in
Refs.~\cite{Lu:1998nu,Claus:1998yw}, and its solution is expressed in a
simpler form if we split $\eta^I$ as
\begin{align}
\eta^I = \eta^I_+ + \eta^I_- \,,
\label{kspinor}
\end{align}
where $\eta^I_\pm$ are defined by
\begin{align}
\eta^I_\pm = P^{IJ}_\pm \eta^J
\end{align}
with the projection operator
\begin{align}
P^{IJ}_\pm =\frac{1}{2} (\delta^{IJ} \pm \Gamma_{0123} \tau_2^{IJ}) \,.
\label{proj}
\end{align}
In this splitting, we see that $\eta^1_\pm$ and $\eta^2_\pm$ are not
independent from each other because
\begin{align}
\eta^2_\pm = \mp \Gamma_{0123} \eta^1_\pm \,.
\label{eta12}
\end{align}
Thus, to avoid this redundancy, it is convenient to define
\begin{align}
\eta_\pm \equiv \eta^1_\pm \,,
\label{etapm}
\end{align}
to which $\eta^1$ and $\eta^2$ are related by
\begin{align}
\eta^1 = \eta_+ + \eta_- \,, \quad
\eta^2 = -\Gamma_{0123} (\eta_+ - \eta_-) \,,
\label{ks12}
\end{align}
as checked from Eqs.~(\ref{kspinor}), (\ref{eta12}) and (\ref{etapm}).
If we now use $\eta_\pm$, then the solution of the Killing spinor equation
is expressed as\footnote{While the complex spinor notation is adopted in
\cite{Claus:1998yw}, we use the real or Majorana-Weyl spinor notation
throughout the paper.}
\begin{align}
\eta_+ (X) &=
  u^{1/2} S (\phi) (\epsilon_+ - x \cdot \Gamma \epsilon_-) \,,
\notag \\
\eta_- (X) &= u^{-1/2} \Gamma_4 S(\phi) \epsilon_-  \,,
\label{kspm}
\end{align}
where $\epsilon_\pm$ are constant spinors,
$x \cdot \Gamma = x^0 \Gamma_0 + x^1 \Gamma_1 +x^2 \Gamma_2 +x^3 \Gamma_3$
(or $x^0 \Gamma_0 + \vec{x} \cdot \vec{\Gamma}$),
and $S(\phi)$ is a spinorial function of five angles of S$^5$ given by
\begin{align}
S(\phi) = \prod^9_{a'=5} \exp
     \left(\frac{1}{2} \phi^{a'-4} \Gamma_{(a'-1) a'} \right) \,.
\label{umat}
\end{align}
We note that, since $\eta^I$ is taken to have positive chirality in
this paper, $\epsilon_+ (\epsilon_-)$ is a positive (negative) chirality
spinor,
\begin{align}
\Gamma^{11} \epsilon_\pm = \pm \epsilon_\pm \,,
\label{epch}
\end{align}
and has sixteen independent free components.\footnote{The definition of
$\Gamma^{11}$ is given in (\ref{gdef}).}

\subsection{Symmetries of D$p$-brane action}

The D$p$-brane action $S_p$ is composed of the Dirac-Born-Infeld (DBI) and the
Wess-Zumino (WZ) parts:
\begin{align}
S_p = S_\text{DBI} + S_\text{WZ} \,,
\label{d3action}
\end{align}
where
\begin{align}
S_\text{DBI} =
    - \int_{\mathcal{M}_{p+1}} d^{p+1} \sigma
      \sqrt{-\det(G_{ij}+\mathcal{F}_{ij})} \,, \quad
S_\text{WZ} = \int_{\mathcal{M}_{p+2}}  H_{p+2} \,,
\end{align}
Here, $\mathcal{M}_{p+1}$ represents the D$p$-brane worldvolume and
$\mathcal{M}_{p+2}$ is a $(p+2)$-dimensional manifold whose boundary  is
identified with $\mathcal{M}_{p+1}$,
that is, $\partial \mathcal{M}_{p+2} = \mathcal{M}_{p+1}$.

In the DBI part, $G_{ij}$ is the pullback of the AdS$_5\times$S$^5$
supergeometry described by the Cartan one-form vector superfield
$L^{\hat{a}}$ onto the worldvolume,
\begin{align}
G_{ij} = L_i^{\hat{a}} L_j^{\hat{b}} \eta_{\hat{a}\hat{b}} \,, \quad
L_i^{\hat{a}} = \partial_i Z^M L_M^{\hat{a}} \,,
\end{align}
where $i,j$ are the worldvolume indices $(i,j=0,1,\dots,p)$.
 $\mathcal{F}_{ij}$ is a combination of the field
strength $F_{ij}$ of the worldvolume gauge field $A_i$
($F_{ij}= \partial_i A_j - \partial_j A_i$) and
the pulled-back background NS-NS two-form superfield $\mathcal{B}$.
In the form notation, $\mathcal{F}$ is given by
\begin{align}
\mathcal{F}= F-\mathcal{B} = dA + 2 i \int^1_0 ds L_s^{\hat{a}} \wedge
    \bar{\Theta}^I \Gamma_{\hat{a}} \tau_3^{IJ} L^J_s \,,
\label{superf}
\end{align}
where $L^I$ is the Cartan one-form spinorial superfield and the
subscript $s$ in the superfields means that the fermionic coordinate
$\Theta$ inside the superfields is replaced by
$\Theta \rightarrow s \Theta$.  In the WZ part, $H_{p+2}$ is the
supersymmetric closed $(p+2)$-form consisting of various combinations of
the Cartan one-form superfields and $\mathcal{F}$.\footnote{For a
comprehensive study on the WZ part, see for example
Ref.~\cite{Sakaguchi:2006pg} where the systematic Chevalley-Eilenberg
cohomology has been used to construct $H_{p+2}$.}

The D$p$-brane action has three manifest symmetries.
Firstly, it is invariant under the worldvolume reparametrization
\begin{align}
\sigma^i \rightarrow \sigma^i - \lambda^i(\sigma) \,,
\label{difft}
\end{align}
where $\lambda^i(\sigma)$ is the local reparametrization parameter.
Under this, the worldvolume fields transform as follows.
\begin{align}
\delta_\lambda \Theta^I = \lambda^i \partial_i \Theta^I \,, \quad
\delta_\lambda X^\mu    = \lambda^i \partial_i X^\mu  \,, \quad
\delta_\lambda A_i = \lambda^j \partial_j A_i + \partial_i \lambda^j A_j \,.
\label{diff1}
\end{align}
We note that both of $\Theta^I$ and $X^\mu$ are scalars from the
worldvolume viewpoint.
Secondly, the action is spacetime supersymmetric under the transformations
\begin{align}
\delta_\eta Z^M L_M^{\hat{a}}
 = 2 i \bar{\eta}^I \Gamma^{\hat{a}} \Theta^I \,, \quad
\delta_\eta Z^M L_M^I
 = \eta^I \,,
\label{stsusyt}
\end{align}
where $\eta^I$ is the Killing spinor of Eq.~(\ref{ks12}) with
Eq.~(\ref{kspm}).  More precisely, the DBI and the WZ parts of the action are
supersymmetric separately.
Actually, supersymmetry is natural because the super coset method
respects the background supersymmetry by construction.
If we expand the spacetime supersymmetry transformation of
Eq.~(\ref{stsusyt}) in terms of $\Theta$, we get
\begin{align}
\delta_\eta \Theta^I &= \eta^I + \mathcal{O}(\Theta^2) \,,
\notag \\
\delta_\eta X^\mu &= - i e^\mu_{\hat{a}} \bar{\Theta}^I \Gamma^{\hat{a}} \eta^I
                     + \mathcal{O}(\Theta^3) \,,
\notag \\
\delta_\eta A_i &= -i e^{\hat{a}}_i \bar{\Theta}^I \Gamma_{\hat{a}}
                     \tau_3^{IJ} \eta^J  + \mathcal{O}(\Theta^3) \,,
\label{susy1}
\end{align}
where $e^\mu_{\hat{a}}$ is the inverse of the zehnbein $e_\mu^{\hat{a}}$
given in Eq.~(\ref{zbein}) and
\begin{align}
e^{\hat{a}}_i = \partial_i X^\mu  e^{\hat{a}}_\mu \,.
\label{pbzb}
\end{align}
The transformation for the worldvolume gauge field $A_i$ is determined
from the invariance of $\mathcal{F}$ of Eq.~(\ref{superf}),
$\delta_\eta \mathcal{F} = 0$\cite{Aganagic:1996pe}.

The last one is the local fermionic $\kappa$ symmetry, which is
in some sense the most important one since it guarantees the
worldvolume supersymmetry after gauge fixing.  The $\kappa$ symmetry
transformation rules are given by
\begin{align}
\delta_\kappa Z^M L_M^{\hat{a}} = 0 \,, \quad
\delta_\kappa Z^M L_M^I
    = \kappa^I \,,
\label{kappat}
\end{align}
where the transformation parameter $\kappa$ satisfies,
for the $\kappa$ symmetric projection $\Gamma^{(p)}$,
\begin{align}
\Gamma^{(p)IJ} \kappa^J = \kappa^I \,.
\label{kappaproj}
\end{align}
The $\kappa$ symmetry projection is basically the pullback of
various gamma matrix products onto the D$p$-brane worldvolume
and, for the type IIB case, its explicit expression \cite{Bergshoeff:1997kr}
is
\begin{align}
\Gamma^{(p)} &= \frac{1}{\sqrt{-\det (G_{ij}+\mathcal{F}_{ij})}}
\sum^{(p+1)/2}_{n=0} \frac{1}{2^n n!} \gamma^{j_1k_1 \cdots j_n k_n}
\mathcal{F}_{j_1k_1} \cdots \mathcal{F}_{j_nk_n} J^{(n)}_{(p)} \,,
\notag \\
J^{(n)}_{(p)} &= \frac{(-1)^n}{(p+1)!}
\epsilon^{i_0 i_1 \cdots i_p}  \gamma_{i_0 i_1 \cdots i_p}
\tau_3^{n+(p-3)/2} \tau_2 \,,
\label{projexp}
\end{align}
where $\gamma_{i_1\cdots i_n} = \gamma_{[i_1} \cdots \gamma_{i_n]}$
and $\gamma_i$ is the pullback of $\Gamma_{\hat{a}}$,
$\gamma_i = L^{\hat{a}}_i \Gamma_{\hat{a}}$.
The important properties of $\Gamma^{(p)}$ are
\begin{align}
\Gamma^{(p)2} =1 \,, \quad
\mbox{Tr}\Gamma^{(p)}=0 \,,
\label{gprop}
\end{align}
and, as is verified with the $\tau$ matrices (\ref{taumat}),
$\Gamma^{(p)}$ can always be put into the form
\begin{align}
\Gamma^{(p)} = \begin{pmatrix}
                     0 & \beta^{(p)}_+ \\ \beta^{(p)}_- & 0
               \end{pmatrix} \,.
\label{beta}
\end{align}
The two blocks $\beta^{(p)}_+$ and $\beta^{(p)}_-$ satisfy
$\beta^{(p)}_+ \beta^{(p)}_- = \beta^{(p)}_- \beta^{(p)}_+ =1$,
which is $\Gamma^{(p)2} =1$ of (\ref{gprop}), and their expressions for
each $p$ will be given in the next section.
If we now write down the $\kappa$ symmetry transformation rules for the
worldvolume fields by expanding Eq.~(\ref{kappat}) in terms of $\Theta$,
then they are
\begin{align}
\delta_\kappa \Theta^I &= \kappa^I + \mathcal{O}(\Theta^2) \,,
\notag \\
\delta_\kappa X^\mu &= i e^\mu_{\hat{a}} \bar{\Theta}^I \Gamma^{\hat{a}}
                   \delta_\kappa \Theta^I + \mathcal{O}(\Theta^3) \,,
\notag \\
\delta_\kappa A_i &= i e^{\hat{a}}_i \bar{\Theta}^I \Gamma_{\hat{a}}
                     \tau_3^{IJ}  \delta_\kappa \Theta^J
                     + \mathcal{O}(\Theta^3) \,,
\label{kappa1}
\end{align}
where the transformation rule for $A_i$ is determined from
$\delta_\kappa \mathcal{F} = -2i L^{\hat{a}} \wedge \bar{L}^I
\Gamma_{\hat{a}} \tau_3^{IJ}  \delta_\kappa \Theta^J$
\cite{Aganagic:1996pe}.\footnote{The expression for
$\delta_\kappa \mathcal{F}$ itself has bee derived in
Ref.~\cite{Metsaev:1998hf}.}

\subsection{Worldvolume supersymmetry}
\label{wvsusy}

If a given configuration or embedding of D$p$-brane in a specific
supersymmetric background preserves some fraction of the background
supersymmetries, then the preserved supersymmetries should be respected on
the D-brane worldvolume theory.  How are they realized and described?
One way to answer this practical question is to follow the
procedure developed in Ref.~\cite{Aganagic:1996pe}.  In this subsection,
following Ref.~\cite{Aganagic:1996pe}, we describe how to identify the
supersymmetry on the worldvolume and give the associated
supersymmetry transformation rules for the worldvolume fields.

For a given D$p$ brane, we first consider its configuration based on the data
of Table \ref{tablebps} and align the worldvolume coordinates with those of
spacetime as
\begin{align}
X^{\ell_i}(\sigma) = \sigma^i \quad (i=0,1,\dots,p) \,,
\label{static}
\end{align}
which is equivalent to specify indices $(\ell_0, \ell_1, \dots \ell_p)$
among ten spacetime coordinates (\ref{10dcoord}).
This is nothing but the static gauge which fixes the worldvolume
reparametrization symmetry (\ref{difft}).  Since the Lorentzian branes are of
our concern, $X^{\ell_0}$ will be always $X^0(=x^0)$, that is, $\ell_0=0$ or
$x^0(\sigma)=\sigma^0$.  The remaining spacetime coordinates transverse to
$X^{\ell_i}$ will be denoted by $X^f$, which describe the transverse
fluctuations of D$p$-brane.  Since the brane may be placed in some
transverse position, it is convenient to split $X^f$ as
\begin{align}
X^f = X_0^f + \tilde{X}^f \,,
\end{align}
where $X_0^f$ denote the constant transverse position of brane and
$\tilde{X}^f$ are the fluctuations around them.
We note that we could consider more general
configurations where $X^f$ depend on $X^{\ell_i}$ as $X^f = X^f(X^{\ell_i})$.
One typical example would be the constant motion along certain transverse
directions: $\partial_0 X^f = \text{constant}$.  In this paper, however, we
will restrict ourselves to purely static configurations and turn off any
worldvolume fluxes

As alluded to in the last subsection, the D$p$ brane has a local worldvolume
symmetry, the $\kappa$ symmetry.  As we do in a theory with local gauge
symmetries, we should fix it properly before doing any actual calculation.
Here, we take the covariant $\kappa$ symmetry fixing condition given by
\begin{align}
\Theta^1 = 0 \,, \quad \Theta^2 = \theta \,,
\label{covgauge}
\end{align}
or $(1+\tau_3)^{IJ}\Theta^J=0$.  This condition is an admissible one because
it is in accord with the criterion for the admissible fixing condition
\cite{Bergshoeff:1997kr}: $\tau_3$ does not commute with $\Gamma^{(p)}$
of the $\kappa$ symmetry projector (\ref{kappaproj}),
$[\tau_3, \Gamma^{(p)}] \neq 0$.  Though the covariant fixing condition
is not so helpful in simplifying the D-brane action, it is convenient
to explore the symmetry structure in a covariant way.\footnote{See for
example Ref.~\cite{Metsaev:2002sg}, where the covariant gauge is adopted
in studying the D3-brane in the plane wave background.}

Having fixed the reparametrization and the $\kappa$ symmetries,  we look at
the spacetime supersymmetry transformation (\ref{susy1}).  Then we easily
see that the transformation violates the gauge-fixing conditions of
Eqs.~(\ref{static}) and (\ref{covgauge}) because $\delta_\eta \Theta^1 \neq 0$
and $\delta_\eta X^{\ell_i} \neq 0$.  One possible way of resolving this
situation is to introduce the compensating $\kappa$ and the worldvolume
reparametrization transformations and define a new transformation $\delta$ as
\begin{align}
\delta \equiv \delta_\eta + \delta_\kappa + \delta_\lambda \,.
\label{newsusy}
\end{align}
The parameters $\kappa$ and $\lambda$ of the compensating transformations are
determined in terms of $\eta^I$ such that the new transformation $\delta$ keeps
the gauge-fixing conditions, that is, $\delta \Theta^1 = 0$ and
$\delta X^{\ell_i} = 0$.  They can be found order by oder in $\theta$ and are,
at the leading order,
\begin{gather}
\kappa^1 = - \eta^1 + \mathcal{O} (\theta^2) \,, \quad
\kappa^2 = - \beta^{(p)}_- \eta^1 + \mathcal{O} (\theta^2) \,,
\notag \\
\lambda^i = i e^{\ell_i}_{\hat{a}}
            \bar{\theta} \Gamma^{\hat{a}} (\eta^2 + \beta^{(p)}_- \eta^1)
            + \mathcal{O} (\theta^3) \,,
\label{para}
\end{gather}
where $\beta^{(p)}_-$ appears due to Eq.~(\ref{beta}).
In this way, we have the transformation $\delta$ consistent with the
gauge-fixing conditions and interpret it as the worldvolume supersymmetry.

Let us now turn to the theory on the D$p$-brane worldvolume.  From the
viewpoint of the worldvolume theory, the static gauge describing the
embedding of the brane can be regarded as the `vacuum'
configuration.  Then, as usual, the supersymmetry preserved by the
`vacuum' is specified by the free components of the supersymmetry parameter
satisfying the equation
$\delta\text{(fermion)}=0$, that is, $\delta \theta = 0$, which we call
the worldvolume Killing spinor equation.  If we rewrite $\delta \theta=0$
by plugging (\ref{diff1}), (\ref{susy1}), (\ref{kappa1}) into
(\ref{newsusy}) with (\ref{para}), it becomes a fairly simple equation,
\begin{align}
0 =& \, \eta^2  - \beta^{(p)}_0 \eta^1 \,,
\label{kse0}
\end{align}
where $\beta^{(p)}_0$ depends on the `vacuum' configuration and is
defined by
\begin{align}
\beta^{(p)}_0 \equiv \left. \beta^{(p)}_-
   \right|_{\theta=0,\,\, A_i=0, \,\, X^f=X^f_0} \,.
\label{beta0}
\end{align}
We would like to note that the worldvolume Killing spinor equation
(\ref{kse0}) is an exact one because $\theta=0$ in the `vacuum'
configuration.

Further evaluation of (\ref{kse0}) using (\ref{ks12}) and (\ref{kspm})
leads us to have
\begin{align}
0 =& -(\Gamma_{0123} + \beta^{(p)}_0) \eta_+
     +(\Gamma_{0123} - \beta^{(p)}_0) \eta_-
\notag \\
  =& - u^{1/2} \Gamma_{0123}(1-\Gamma_{0123} \beta^{(p)}_0)
     S_0 (\phi) (\epsilon_+ - x \cdot \Gamma \epsilon_-)
\notag \\
   & + u^{-1/2} \Gamma_{0123}(1+\Gamma_{0123} \beta^{(p)}_0)
     \Gamma_4 S_0 (\phi) \epsilon_- \,,
\label{kse}
\end{align}
where some coordinates among $u$ and $x^{1,2,3}$ are understood to be
constants if they are transverse directions, and $S_0 (\phi)$ means
$S(\phi)$ of (\ref{umat}) in which the angular directions included in
$X^f$ are set to constant values.
In the investigation of the worldvolume supersymmetry structure of a
`vacuum' configuration, the first step is to check if the square of
$\Gamma_{0123} \beta^{(p)}_0$ in Eq.~(\ref{kse}) is equal to one,
$\left( \Gamma_{0123} \beta^{(p)}_0 \right)^2 = 1$.
If it is the case, $\Gamma_{0123} \beta^{(p)}_0$ has eigenvalues of
$\pm 1$.  This means that $1 \pm \Gamma_{0123} \beta^{(p)}_0$ play the role
of projection operators and give us the possibility of identifying which
components of $\epsilon_\pm$ are free.  The next step is to send
$\Gamma_{0123} \beta^{(p)}_0$ to the right of $S_0 (\phi)$, which is
done by evaluating $S_0^{-1} (\phi) \Gamma_{0123} \beta^{(p)}_0 S_0 (\phi)$
with repeated use of the following identity
\begin{align}
e^{-\frac{1}{2} \phi \Gamma_{a(a+1)}} \Gamma_a
e^{\frac{1}{2} \phi \Gamma_{a(a+1)}}
= \Gamma_a \cos \phi + \Gamma_{a+1} \sin \phi \,.
\label{iden}
\end{align}
Generically, the resulting expression is not of the form of projection operator
but a sum of many gamma matrix products with coefficients composed of
trigonometric functions.  However, as we will see in the next section,
it becomes a projection operator for some special values of the transverse
angular coordinates and can be used to pick out the free components among
$\epsilon_\pm$.  This is interesting in a sense that the transverse position
is determined by insisting on the supersymmetry in the D-brane
worldvolume theory without resort to the equations of
motion.\footnote{Strictly speaking, in some cases, we should also impose
the non-degeneracy condition for the induced worldvolume metric.}

After identifying the supersymmetry preserved on the D-brane worldvolume,
we can read off how the transformation $\delta$ acts on the remaining
worldvolume fields.  Let us denote the worldvolume fields collectively as
\begin{align}
\Phi = (\tilde{X}^f, A_i, \theta) \,.
\end{align}
Its transformation $\delta \Phi$ can be written as an expansion in terms
of the power of $\Phi$.  If we consider the terms up to linear order in
$\Phi$, then the transformation rules for the worldvolume fields are
\begin{align}
\delta \tilde{X}^f &= - i e^f_{\hat{a}} \bar{\theta} \Gamma^{\hat{a}}
                    (\hat{\eta}^2 + \beta^{(p)}_- \hat{\eta}^1)
                 + \mathcal{O} (\Phi^2) \,,
\notag \\
\delta A_i &=   i e^{\hat{a}}_i \bar{\theta} \Gamma_{\hat{a}}
                    (\hat{\eta}^2 + \beta^{(p)}_- \hat{\eta}^1)
                 + \mathcal{O} (\Phi^2) \,,
\notag \\
\delta \theta &= \hat{\eta}^2 - \beta^{(p)}_- \hat{\eta}^1
                 + \mathcal{O} (\Phi^2)  \,,
\label{susyt}
\end{align}
where $\hat{\eta}^{1,2}$ are $\eta^{1,2}$ containing only the free components
of $\epsilon_\pm$ picked out from the worldvolume Killing spinor equation
(\ref{kse}) and we have omitted the compensating worldvolume
reparametrization transformation because it begins with the terms
quadratic order in $\Phi$.  In these transformation rules, $\beta^{(p)}_-$
is also expanded as
\begin{align}
\beta^{(p)}_- = \beta^{(p)}_0 + \beta^{(p)}_1 + \mathcal{O} (\Phi^2) \,,
\label{betam}
\end{align}
where $\beta^{(p)}_0$ is defined in (\ref{beta0}) and $\beta^{(p)}_1$
is the collection of terms linear order in $\Phi$.

In the process of calculation leading to (\ref{susyt}), we will encounter
many trigonometric functions.  Thus, for notational simplicity, we would like
to define the following quantities before moving on to the next section.
\begin{align}
s_\alpha \equiv \sin \phi^\alpha \,, \quad c_\alpha \equiv \cos \phi^\alpha \,,
\notag \\
\mathring{s}_\alpha \equiv \sin \phi_0^\alpha \,, \quad
\mathring{c}_\alpha \equiv \cos \phi_0^\alpha \,,
\label{trig}
\end{align}

\section{AdS branes}
\label{adsbrane}

When the radial direction of the AdS space $u$ is one of the worldvolume
directions for a given brane configuration, the brane is usually called
the AdS brane since the induced metric on the worldvolume contains the
AdS space.  If we take a look at the Table \ref{tablebps} and consider
the Lorentzian branes in which the time $x^0$ is always a worldvolume
direction, we see that there are six types of AdS brane configurations.
In this section, for each of them, following the procedure
outlined in Sec.~\ref{wvsusy}, we investigate the supersymmetry
realized on the worldvolume and give the worldvolume supersymmetry
transformation rules for the worldvolume fields .  We note that the following
subsections and subsubsections are self-contained and completely independent
from each other.

\subsection{D1}

The D1-brane configuration (2,0) of Table \ref{tablebps} leads us to take
the static gauge as
\begin{align}
x^0(\sigma) = \sigma^0 \,, \quad
u(\sigma) = \sigma^1 \,,
\label{20static}
\end{align}
or $(\ell_0, \ell_1)=(0,4)$ in Eq.~(\ref{static}),
which corresponds to the AdS$_2$ brane.  The coordinates transverse to this
configuration are then
\begin{align}
X^f = \{ x^1, x^2, x^3, \phi^1, \dots , \phi^5 \} \,.
\end{align}

In order to identify which part of the spacetime supersymmetry is
preserved on the worldvolume of AdS$_2$ brane, we should solve the
worldvolume Killing spinor equation (\ref{kse0}).  What is necessary
to do this is $\beta^{(1)}_\pm$, which  is read off from
Eqs.~(\ref{projexp}) and (\ref{beta}) as
\begin{align}
\beta^{(1)}_\pm =
    \frac{\epsilon^{i_1 i_2}}{2\sqrt{-\det (G_{ij} + \mathcal{F}_{ij})}}
    \bigg(
             \gamma_{i_1 i_2} \pm \mathcal{F}_{i_1 i_2}
    \bigg) \,.
\end{align}
According to (\ref{beta0}), we then see that this expression leads to
\begin{align}
\beta^{(1)}_0 = -\Gamma_{04}
\end{align}
in the static gauge (\ref{20static}).  Now
$\left( \Gamma_{0123} \beta^{(1)}_0 \right)^2 = 1$ obviously and thus
$1 \pm \Gamma_{0123} \beta^{(1)}_0$ in the worldvolume Killing spinor
equation (\ref{kse}) with $p=1$ play the role of projection operators.
Having the projection operators, the next step described
in Sec.~\ref{wvsusy} is to send $\Gamma_{0123} \beta^{(1)}_0$ to the
right of $S_0(\phi)$ in (\ref{kse}).  If we denote the resulting
expression as $\tilde{\Gamma}$, we get the relation
$\Gamma_{0123} \beta^{(1)}_0 S_0(\phi)= S_0(\phi) \tilde{\Gamma}$.
With the fact that
$\Gamma_{0123} \beta^{(1)}_0 = - \Gamma_{1234}$, $\tilde{\Gamma}$
is evaluated by repeated use of the identity (\ref{iden}) as follows:
\begin{align}
-\tilde{\Gamma}
&= S_0^{-1} (\phi) \Gamma_{1234} S_0 (\phi)
\notag \\
&=
    \mathring{c}_1 \Gamma_{1234}
  + \mathring{s}_1 \mathring{c}_2 \Gamma_{1235}
  + \mathring{s}_1 \mathring{s}_2 \mathring{c}_3 \Gamma_{1236}
  + \mathring{s}_1 \mathring{s}_2 \mathring{s}_3 \mathring{c}_4 \Gamma_{1237}
\notag \\
& \phantom{=\,}
  + \mathring{s}_1 \mathring{s}_2 \mathring{s}_3 \mathring{s}_4
        \mathring{c}_5 \Gamma_{1238}
  + \mathring{s}_1 \mathring{s}_2 \mathring{s}_3 \mathring{s}_4
        \mathring{s}_5 \Gamma_{1239} \,,
\end{align}
where we have used the definitions of Eq.~(\ref{trig}).  However,
this shows clearly that $1\pm \tilde{\Gamma}$ do not have the form of
projection operators.  At this stage, we are required to fix the
transverse angular position properly in such a way that makes them have
the desired form.
Although there are various possibilities, we fix the angular
position to be $\phi_0^\alpha = \frac{\pi}{2}$ ($\alpha=1,\dots,5$) for
simplicity,
since all the points on S$^5$ are equivalent and the AdS$_2$ brane is
a point on S$^5$.  For this angular position,
$\tilde{\Gamma} = -\Gamma_{1239}$.  If we now split
$\epsilon_\pm$ according to the eigenvalues of $\Gamma_{1234}$ as
\begin{align}
\Gamma_{1239} \epsilon_{+\pm} = \pm \epsilon_{+\pm} \,, \quad
\Gamma_{1239} \epsilon_{-\pm} = \pm \epsilon_{-\pm} \,,
\end{align}
then the worldvolume Killing spinor equation (\ref{kse}) becomes
\begin{align}
0 = - 2 u^{1/2} \Gamma_{0123} S_0(\phi)
       (\epsilon_{++} - x^0 \Gamma_0 \epsilon_{-+}
         - \vec{x}_0 \cdot \vec{\Gamma} \epsilon_{--} )
    + 2 u^{-1/2} \Gamma_{0123} \Gamma_4 S_0(\phi) \epsilon_{-+} \,,
\end{align}
with
\begin{align}
S_0(\phi) =  S(\phi)|_{\phi^1,\dots,\phi^5=\pi/2} \,.
\end{align}
The solution of this equation is readily found to be
\begin{align}
\epsilon_{-+} = 0 \,, \quad
\epsilon_{++} = \vec{x}_0 \cdot \vec{\Gamma} \epsilon_{--} \,.
\label{20fix}
\end{align}
Since other components except for those of (\ref{20fix}) are
undetermined, we conclude that the supersymmetry preserved on the AdS$_2$
brane is characterized by
\begin{align}
\epsilon_{+-} \,, \quad \epsilon_{--} \,,
\label{20ep}
\end{align}
each of which has eight free components and we have sixteen
supersymmetries (1/2-BPS) in total.

Having identified the worldvolume supersymmetries (\ref{20ep}), it is
straightforward to obtain the supersymmetry transformation rules for the
worldvolume fields according to (\ref{susyt}) and (\ref{betam}).
Firstly, the scalar fields corresponding to the transverse fluctuations
are found to transform as
\begin{align}
\delta \vec{\tilde{x}} &=
    2i u^{-1} \bar{\theta} \vec{\Gamma} \Gamma_{0123}
    (\hat{\eta}_+ - \hat{\eta}_-) + \dots \,,
\notag \\
\delta \tilde{\phi}^{\alpha} &=
     2i \bar{\theta} \Gamma_{\alpha+4} \Gamma_{0123}
      (\hat{\eta}_+ - \hat{\eta}_-) + \dots \,,
\end{align}
where $\alpha = 1,\dots,5$ and
\begin{align}
\hat{\eta}_+ &=
    u^{1/2} S_0(\phi)
    \left(
        \epsilon_{+-} - x^0 \Gamma_0 \epsilon_{--}
    \right) \,,
\notag \\
\hat{\eta}_- &=
    u^{-1/2} \Gamma_4 S_0(\phi) \epsilon_{--} \,.
\end{align}
As for the worldvolume gauge field, the transformation rule
is obtained as
\begin{align}
\delta A_0 &=
    -2i u \bar{\theta} \Gamma_0 \Gamma_{0123}
    (\hat{\eta}_+ - \hat{\eta}_-) + \dots \,,
\notag \\
\delta A_u &=
    -2i u^{-1} \bar{\theta} \Gamma_4 \Gamma_{0123}
    (\hat{\eta}_+ - \hat{\eta}_-) + \dots \,.
\end{align}
Finally, we get the transformation rule for the fermionic field as
\begin{align}
\delta \theta &=
    - 2 u \vec{\tilde{x}} \cdot \vec{\Gamma} \Gamma_{01234} \hat{\eta}_-
    + \tilde{\phi}^\alpha \Gamma_{\alpha+4} \Gamma_{01234}
        (\hat{\eta}_+ + \hat{\eta}_-)
    - \beta^{(1)}_1 (\hat{\eta}_+ + \hat{\eta}_-) + \dots \,,
\end{align}
where $\alpha=1,\dots,5$ and
\begin{align}
\beta^{(1)}_1 &=
       \left(
           \frac{1}{u} \Gamma^0 \partial_0 \tilde{X}^f
           + u \Gamma^4 \partial_u \tilde{X}^f
       \right) e_f^{\hat{a}} \Gamma_{0 4 \hat{a}}
    +  F_{0 u} \,.
\end{align}

\subsection{D3}
\label{31config}

The D3-brane configuration (3,1) of Table \ref{tablebps} leads us to take
the static gauge as
\begin{align}
x^{0,1}(\sigma) = \sigma^{0,1} \,, \quad
u(\sigma) = \sigma^2 \,, \quad
\phi^5 (\sigma) = \sigma^3 \,,
\label{31static}
\end{align}
or $(\ell_0, \ell_1, \ell_2, \ell_3)=(0,1,4,9)$ in Eq.~(\ref{static}),
which corresponds to the AdS$_3 \times$S$^1$ brane.  The coordinates transverse
to this configuration are then
\begin{align}
X^f = \{ x^2, x^3, \phi^1, \dots , \phi^4 \} \,.
\end{align}

In order to identify which part of the spacetime supersymmetry is
preserved on the worldvolume of AdS$_3 \times$S$^1$ brane, we should solve
the worldvolume Killing spinor equation (\ref{kse0}).  What is necessary
to do this is $\beta^{(3)}_\pm$, which  is read off from
Eqs.~(\ref{projexp}) and (\ref{beta}) as
\begin{align}
\beta^{(3)}_\pm =
    \frac{\epsilon^{i_1 \cdots i_4}}{\sqrt{-\det (G_{ij}+\mathcal{F}_{ij})}}
    \bigg(
         \pm \frac{1}{4!} \gamma_{i_1 \cdots i_4}
          +  \frac{1}{4}  \gamma_{i_1 i_2} \mathcal{F}_{i_3 i_4}
         \pm \frac{1}{8}  \mathcal{F}_{i_1 i_2} \mathcal{F}_{i_3 i_4}
    \bigg) \,.
\end{align}
According to (\ref{beta0}), we then see that this expression leads to
\begin{align}
\beta^{(3)}_0 = \Gamma_{0149}
\end{align}
in that static gauge (\ref{31static}).  Now
$\left( \Gamma_{0123} \beta^{(3)}_0 \right)^2 = 1$ obviously and thus
$1 \pm \Gamma_{0123} \beta^{(3)}_0$ in the worldvolume Killing spinor
equation (\ref{kse}) with $p=3$ play the role of projection operators.
Having the projection operators, the next step described
in Sec.~\ref{wvsusy} is to send $\Gamma_{0123} \beta^{(3)}_0$ to the
right of $S_0(\phi)$ in (\ref{kse}).  If we denote the resulting
expression as $\tilde{\Gamma}$, we get the relation
$\Gamma_{0123} \beta^{(3)}_0 S_0(\phi)= S_0(\phi) \tilde{\Gamma}$.
With the fact that
$\Gamma_{0123} \beta^{(3)}_0 = \Gamma_{2349}$, $\tilde{\Gamma}$
is evaluated by repeated use of the identity (\ref{iden}) as follows:
\begin{align}
\tilde{\Gamma}
&= S_0^{-1} (\phi) \Gamma_{2349} S_0 (\phi)
\notag \\
&=
    \mathring{c}_1 \Gamma_{2349}
    + \mathring{s}_1 \mathring{c}_2 \Gamma_{2359}
    + \mathring{s}_1 \mathring{s}_2 \mathring{c}_3 \Gamma_{2369}
    - \mathring{s}_1 \mathring{s}_2 \mathring{s}_3
      \mathring{c}_4 s_5 \Gamma_{2378}
\notag \\
& \phantom{=\,}
    + \mathring{s}_1 \mathring{s}_2 \mathring{s}_3 \mathring{c}_4
        c_5 \Gamma_{2379}
    + \mathring{s}_1 \mathring{s}_2 \mathring{s}_3
        \mathring{s}_4 \Gamma_{2389} \,,
\end{align}
where we have used the definitions of Eq.~(\ref{trig}).  However,
this shows clearly that $1\pm \tilde{\Gamma}$ do not have the form of
projection operators.  One can make them have the desired form by fixing
the transverse angular position, and realize that there are four
possible choices which are (i) $\phi^1_0 =0$, $\phi_0^{2,3,4} =$arbitrary,
(ii) $\phi^1_0 = \frac{\pi}{2}$, $\phi^2_0 = 0$, $\phi^{3,4}_0 =$arbitrary,
(iii) $\phi^{1,2}_0 = \frac{\pi}{2}$, $\phi^3_0 = 0$,
$\phi^4_0 =$arbitrary, (iv) $\phi^{1,2,3,4}_0 = \frac{\pi}{2}$.  Except
for the last one, the first three choices lead to the singular or
degenerate induced worldvolume metric.  Thus, if one wishes to have
a regular theory on the worldvolume, the last choice is quite natural and
hence $\tilde{\Gamma} = \Gamma_{2389}$.
If we now split
$\epsilon_\pm$ according to the eigenvalues of $\Gamma_{2389}$ as
\begin{align}
\Gamma_{2389} \epsilon_{+\pm} = \pm \epsilon_{+\pm} \,, \quad
\Gamma_{2389} \epsilon_{-\pm} = \pm \epsilon_{-\pm} \,,
\end{align}
then the worldvolume Killing spinor equation (\ref{kse}) becomes
\begin{align}
0 =& - 2 u^{1/2} \Gamma_{0123} S_0 (\phi)
       \left[
           \epsilon_{+-} - (x^0 \Gamma_0 + x^1 \Gamma_1)\epsilon_{--}
           - (x^2_0 \Gamma_2 + x^3_0 \Gamma_3) \epsilon_{-+}
       \right]
\notag \\
 &
    + 2 u^{-1/2} \Gamma_{0123} \Gamma_4 S_0 (\phi) \epsilon_{--} \,,
\end{align}
with
\begin{align}
S_0(\phi) =  S(\phi)|_{\phi^1,\dots,\phi^4=\pi/2} \,.
\end{align}
The solution of this equation is readily found to be
\begin{align}
\epsilon_{--} = 0 \,, \quad
\epsilon_{+-} =  (x^2_0 \Gamma_2 + x^3_0 \Gamma_3) \epsilon_{-+} \,.
\label{31fix}
\end{align}
Since other components except for those of (\ref{31fix}) are
undetermined, we conclude that the supersymmetry preserved on the
AdS$_3 \times$S$^1$ brane is characterized by
\begin{align}
\epsilon_{++} \,, \quad \epsilon_{-+} \,,
\label{31ep}
\end{align}
each of which has eight free components and we have sixteen
supersymmetries (1/2-BPS) in total.

Having identified the worldvolume supersymmetries (\ref{31ep}), it is
straightforward to obtain the supersymmetry transformation rules for the
worldvolume fields according to (\ref{susyt}) and (\ref{betam}).
Firstly, the scalar fields corresponding to the transverse fluctuations
are found to transform as
\begin{align}
\delta \tilde{x}^{2,3} &=
    2i u^{-1} \bar{\theta} \Gamma^{2,3} \Gamma_{0123}
    (\hat{\eta}_+ - \hat{\eta}_-) + \dots \,,
\notag \\
\delta \tilde{\phi}^{1,2,3,4} &=
     2i \bar{\theta} \Gamma^{5,6,7,8} \Gamma_{0123}
      (\hat{\eta}_+ - \hat{\eta}_-) + \dots \,,
\label{xtransf}
\end{align}
where
\begin{align}
\hat{\eta}_+ &=
    u^{1/2} S_0 (\phi)
    \left[
        \epsilon_{++}
        - (x^0 \Gamma_0 + x^1 \Gamma_1 ) \epsilon_{-+}
    \right] \,,
\notag \\
\hat{\eta}_- &=
    u^{-1/2} \Gamma_4 S_0 (\phi) \epsilon_{-+} \,.
\label{d3eta}
\end{align}
As for the worldvolume gauge field, the transformation rule
is obtained as
\begin{align}
\delta A_{0,1} &=
    -2i u \bar{\theta} \Gamma_{0,1} \Gamma_{0123}
    (\hat{\eta}_+ - \hat{\eta}_-) + \dots \,,
\notag \\
\delta A_u &=
    -2i u^{-1} \bar{\theta} \Gamma_4 \Gamma_{0123}
    (\hat{\eta}_+ - \hat{\eta}_-) + \dots \,,
\notag \\
\delta A_{\phi^5} &=
    -2i \bar{\theta} \Gamma_9 \Gamma_{0123}
    (\hat{\eta}_+ - \hat{\eta}_-) + \dots \,,
\label{gtransf}
\end{align}
Finally, we get the transformation rule for the fermionic field as
\begin{align}
\delta \theta &=
    - 2 u ( \tilde{x}^2 \Gamma_2 + \tilde{x}^3 \Gamma_3)
        \Gamma_{01234} \hat{\eta}_-
    + \tilde{\phi}^\alpha \Gamma_{\alpha+4} \Gamma_{01234}
        (\hat{\eta}_+ + \hat{\eta}_-)
    - \beta^{(3)}_1 (\hat{\eta}_+ + \hat{\eta}_-) + \dots \,,
\label{stransf}
\end{align}
where $\alpha = 1,2,3,4$ and
\begin{align}
\beta^{(3)}_1 =& \,
    -  \left(
           \frac{1}{u} \Gamma^0 \partial_0 \tilde{X}^f
           + \frac{1}{u} \Gamma^1 \partial_1 \tilde{X}^f
           + u \Gamma^4 \partial_u \tilde{X}^f
           + \Gamma^9 \partial_{\phi^5} \tilde{X}^f
       \right) e_f^{\hat{a}} \Gamma_{0149 \hat{a}}
\notag \\
 &  \,
    - u \Gamma_{01} F_{u\phi^5} + \frac{1}{u} \Gamma_{04} F_{1\phi^5}
    -  \Gamma_{09} F_{1u} - \frac{1}{u} \Gamma_{14} F_{0\phi^5}
    +  \Gamma_{19} F_{0u} - \frac{1}{u^2} \Gamma_{49} F_{01} \,.
\label{beta31}
\end{align}

One may wonder if the D-brane configuration considered above is
stable since S$^1$ in S$^5$ is not a topological cycle. The similar
problem was worked out by \cite{Karch:2000gx}.
The scalar mode corresponding to slipping off the S$^1$ in S$^5$
satisfies Breitenlohner-Friedmann bound, hence
it does not lead to the instability. Also the above D-brane
configuration satisfies the so called generalized calibration,
which is the condition for the supersymmetric cycle of D-branes to
satisfy on the general supergravity background with various fluxes.
\cite{Cascales:2004qp}. These remarks hold as well for other D-brane
configurations in subsequent subsections.

\subsection{D5}

In this subsection, we are led to consider two kinds of D5-brane
configurations.  The common content for them is $\beta^{(5)}_\pm$
appearing in the $\kappa$ symmetry projection $\Gamma^{(5)}$, which is
read off from Eqs.~(\ref{projexp}) and (\ref{beta}) as
\begin{align}
\beta^{(5)}_\pm =
  & \, \frac{\epsilon^{i_1 \cdots i_6}}{\sqrt{-\det (G_{ij}+\mathcal{F}_{ij})}}
    \bigg(
              \frac{1}{6!} \gamma_{i_1 \cdots i_6}
          \pm \frac{1}{48} \gamma_{i_1 \cdots i_4} \mathcal{F}_{i_5 i_6}
          +   \frac{1}{16} \gamma_{i_1 i_2}
                           \mathcal{F}_{i_3 i_4} \mathcal{F}_{i_5 i_6}
\notag \\
  &       \pm \frac{1}{48} \mathcal{F}_{i_1 i_2} \mathcal{F}_{i_3 i_4}
              \mathcal{F}_{i_5 i_6}
    \bigg) \,.
\label{b5}
\end{align}

\subsubsection{$(4, 2)$-brane}
\label{42config}

The D5-brane configuration (4,2) of Table \ref{tablebps} leads us to take
the static gauge as
\begin{align}
x^{0,1,2}(\sigma) = \sigma^{0,1,2} \,, \quad
u(\sigma) = \sigma^3 \,, \quad
\phi^{4,5} (\sigma) = \sigma^{4,5} \,,
\label{42static}
\end{align}
or $(\ell_0, \ell_1, \ell_2, \ell_3, \ell_4, \ell_5)=(0,1,2,4,8,9)$ in Eq.~(\ref{static}),
which corresponds to the AdS$_4 \times$S$^2$ brane.  The coordinates transverse
to this configuration are then
\begin{align}
X^f = \{ x^3, \phi^1, \phi^2 , \phi^3 \} \,.
\end{align}

In the static gauge (\ref{42static}), $\beta^{(5)}_-$ of (\ref{b5}) becomes,
\begin{align}
\beta^{(5)}_0 = -\Gamma_{012489} \,,
\end{align}
according to (\ref{beta0}).  Now
$\left( \Gamma_{0123} \beta^{(5)}_0 \right)^2 = 1$ obviously and thus
$1 \pm \Gamma_{0123} \beta^{(5)}_0$ in the worldvolume Killing spinor
equation (\ref{kse}) with $p=5$ play the role of projection operators.
Having the projection operators, the next step described
in Sec.~\ref{wvsusy} is to send $\Gamma_{0123} \beta^{(5)}_0$ to the
right of $S_0(\phi)$ in (\ref{kse}).  If we denote the resulting
expression as $\tilde{\Gamma}$, we get the relation
$\Gamma_{0123} \beta^{(5)}_0 S_0(\phi)= S_0(\phi) \tilde{\Gamma}$.
With the fact that
$\Gamma_{0123} \beta^{(5)}_0 = \Gamma_{3489}$, $\tilde{\Gamma}$
is evaluated by repeated use of the identity (\ref{iden}) as follows:
\begin{align}
\tilde{\Gamma}
&= S_0^{-1} (\phi) \Gamma_{3489} S_0 (\phi)
\notag \\
&=  \mathring{c}_1 c_4 \Gamma_{3489}
  + \mathring{c}_1 s_4 s_5 \Gamma_{3478}
  - \mathring{c}_1 s_4 c_5 \Gamma_{3479}
\notag \\
& \phantom{=\,}
  + \mathring{s}_1 \mathring{c}_2 c_4 \Gamma_{3589}
  + \mathring{s}_1 \mathring{c}_2 s_4 s_5 \Gamma_{3578}
  - \mathring{s}_1 \mathring{c}_2 s_4 c_5 \Gamma_{3579}
\notag \\
& \phantom{=\,}
  + \mathring{s}_1 \mathring{s}_2 \mathring{c}_3 c_4 \Gamma_{3689}
  + \mathring{s}_1 \mathring{s}_2 \mathring{c}_3 s_4 s_5 \Gamma_{3678}
  - \mathring{s}_1 \mathring{s}_2 \mathring{c}_3 s_4 c_5 \Gamma_{3679}
\notag \\
& \phantom{=\,}
  + \mathring{s}_1 \mathring{s}_2 \mathring{s}_3 \Gamma_{3789} \,,
\end{align}
where we have used the definitions of Eq.~(\ref{trig}).
This shows clearly that $1\pm \tilde{\Gamma}$ do not have the form of
projection operators.  One can make them have the desired form by fixing
the transverse angular position, and realize that there is a unique
choice of $\phi^{1,2,3}_0 = \frac{\pi}{2}$ which leads to
$\tilde{\Gamma} = \Gamma_{3789}$.
If we now split
$\epsilon_\pm$ according to the eigenvalues of $\Gamma_{3789}$ as
\begin{align}
\Gamma_{3789} \epsilon_{+\pm} = \pm \epsilon_{+\pm} \,, \quad
\Gamma_{3789} \epsilon_{-\pm} = \pm \epsilon_{-\pm} \,,
\end{align}
then the worldvolume Killing spinor equation (\ref{kse}) becomes
\begin{align}
0 =& - 2 u^{1/2} \Gamma_{0123} S_0 (\phi)
       \left[
           \epsilon_{+-}
           - (x^0 \Gamma_0 + x^1 \Gamma_1 + x^2 \Gamma_2)\epsilon_{--}
           - x^3_0 \Gamma_3 \epsilon_{-+}
       \right]
\notag \\
 &
    + 2 u^{-1/2} \Gamma_{0123} \Gamma_4 S_0 (\phi) \epsilon_{--} \,,
\end{align}
with
\begin{align}
S_0(\phi) =  S(\phi)|_{\phi^1,\phi^2,\phi^3=\pi/2} \,.
\end{align}
The solution of this equation is readily found to be
\begin{align}
\epsilon_{--} = 0 \,, \quad
\epsilon_{+-} = x^3_0 \Gamma_3 \epsilon_{-+} \,.
\label{42fix}
\end{align}
Since other components except for those of (\ref{42fix}) are
undetermined, we conclude that the supersymmetry preserved on the
AdS$_4 \times$S$^2$ brane is characterized by
\begin{align}
\epsilon_{++} \,, \quad \epsilon_{-+} \,,
\label{42ep}
\end{align}
each of which has eight free components and we have sixteen
supersymmetries (1/2-BPS) in total.

Having identified the worldvolume supersymmetries (\ref{42ep}), it is
straightforward to obtain the supersymmetry transformation rules for the
worldvolume fields according to (\ref{susyt}) and (\ref{betam}).
Firstly, the scalar fields corresponding to the transverse fluctuations
are found to transform as
\begin{align}
\delta \tilde{x}^3 &=
    2i u^{-1} \bar{\theta} \Gamma^3 \Gamma_{0123}
    (\hat{\eta}_+ - \hat{\eta}_-) + \dots \,,
\notag \\
\delta \tilde{\phi}^{1,2,3} &=
     2i \bar{\theta} \Gamma^{5,6,7} \Gamma_{0123}
      (\hat{\eta}_+ - \hat{\eta}_-) + \dots \,.
\end{align}
where
\begin{align}
\hat{\eta}_+ &=
    u^{1/2} S_0(\phi)
    \left[
        \epsilon_{++}
        - (x^0 \Gamma_0 + x^1 \Gamma_1
           + x^2 \Gamma_2 ) \epsilon_{-+}
    \right] \,,
\notag \\
\hat{\eta}_- &=
    u^{-1/2} \Gamma_4 S_0(\phi) \epsilon_{-+} \,.
\end{align}
As for the worldvolume gauge field, we obtain
\begin{align}
\delta A_{0,1,2} &=
    -2i u \bar{\theta} \Gamma_{0,1,2} \Gamma_{0123}
    (\hat{\eta}_+ - \hat{\eta}_-) + \dots \,,
\notag \\
\delta A_u &=
    -2i u^{-1} \bar{\theta} \Gamma_4 \Gamma_{0123}
     (\hat{\eta}_+ - \hat{\eta}_-) + \dots \,,
\notag \\
\delta A_{\phi^4} &=
    -2i \bar{\theta} \Gamma_8 \Gamma_{0123}
   (\hat{\eta}_+ - \hat{\eta}_-) + \dots \,.
\notag \\
\delta A_{\phi^5} &=
    -2i s_4 \bar{\theta} \Gamma_9 \Gamma_{0123}
   (\hat{\eta}_+ - \hat{\eta}_-) + \dots \,.
\end{align}
Finally, we get the transformation rule for the fermionic field as
\begin{align}
\delta \theta &=
     - 2 u \tilde{x}^3 \Gamma_3 \Gamma_{01234} \hat{\eta}_-
     + \tilde{\phi}^\alpha \Gamma_{\alpha+4} \Gamma_{01234}
        (\hat{\eta}_+ + \hat{\eta}_-)
    - \beta^{(5)}_1 (\hat{\eta}_+ + \hat{\eta}_-) + \dots
\end{align}
where $\alpha=1,2,3$ and
\begin{align}
\beta^{(5)}_1 =& \,
       \bigg(
           \frac{1}{u} \Gamma^0 \partial_0 \tilde{X}^f
           + \frac{1}{u} \Gamma^1 \partial_1 \tilde{X}^f
           + \frac{1}{u} \Gamma^2 \partial_2 \tilde{X}^f
           + u \Gamma^4 \partial_u \tilde{X}^f
\notag \\
 & \,
           + \Gamma^8 \partial_{\phi^4} \tilde{X}^f
           + \frac{1}{s_4} \Gamma^9 \partial_{\phi^5} \tilde{X}^f
       \bigg) e_f^{\hat{a}} \Gamma_{012489 \hat{a}}
\notag \\
 &  \,
    - \frac{1}{48 u^2 s_4} \epsilon^{i_0\cdots i_3 i_4 i_5}
    \left(
        e^{\hat{a}_0}_{\ell_{i_0}} \cdots e^{\hat{a}_3}_{\ell_{i_3}}
    \right) \bigg|_{\phi^1,\phi^2,\phi^3=\pi/2}
    \Gamma_{\hat{a}_0 \cdots \hat{a}_3}
    F_{i_4 i_5} \,.
\end{align}

\subsubsection{$(2, 4)$-brane}

The D5-brane configuration (2,4) of Table \ref{tablebps} leads us to take
the static gauge as
\begin{align}
x^0 (\sigma) = \sigma^0 \,, \quad
u(\sigma) = \sigma^1 \,, \quad
\phi^{2,3,4,5} (\sigma) = \sigma^{2,3,4,5} \,,
\label{24static}
\end{align}
or $(\ell_0, \ell_1, \ell_2, \ell_3, \ell_4, \ell_5)=(0,4,6,7,8,9)$ in Eq.~(\ref{static}), which corresponds to the AdS$_2 \times$S$^4$ brane.
The coordinates transverse to this configuration are then
\begin{align}
X^f = \{ x^1, x^2, x^3, \phi^1 \}
\end{align}

In the static gauge (\ref{24static}), $\beta^{(5)}_-$ of (\ref{b5}) becomes,
\begin{align}
\beta^{(5)}_0 = -\Gamma_{046789} \,,
\end{align}
according to (\ref{beta0}).  Now
$\left( \Gamma_{0123} \beta^{(5)}_0 \right)^2 = 1$ obviously and thus
$1 \pm \Gamma_{0123} \beta^{(5)}_0$ in the worldvolume Killing spinor
equation (\ref{kse}) with $p=5$ play the role of projection operators.
Having the projection operators, the next step described
in Sec.~\ref{wvsusy} is to send $\Gamma_{0123} \beta^{(5)}_0$ to the
right of $S_0(\phi)$ in (\ref{kse}).  If we denote the resulting
expression as $\tilde{\Gamma}$, we get the relation
$\Gamma_{0123} \beta^{(5)}_0 S_0(\phi)= S_0(\phi) \tilde{\Gamma}$.
With the fact that
$\Gamma_{0123} \beta^{(5)}_0 = -\Gamma_{12346789} = -\Gamma^{05} \Gamma^{11}$
where $\Gamma^{11}$ of (\ref{gdef}) has been used, $\tilde{\Gamma}$
is evaluated by repeated use of the identity (\ref{iden}) as follows:
\begin{align}
\tilde{\Gamma}
&= S_0^{-1} (\phi) \Gamma^{05}\Gamma^{11} S_0 (\phi)
\notag \\
&=
   \big( - \mathring{s}_1 \Gamma^{04}
        + \mathring{c}_1 c_2 \Gamma^{05}
        + \mathring{c}_1 s_2 s_3 \Gamma^{06}
        +\mathring{c}_1 s_2 s_3 c_4 \Gamma^{07}
\notag \\
&\phantom{=\,}
        + \mathring{c}_1 s_2 s_3 s_4 c_5 \Gamma^{08}
        + \mathring{c}_1 s_2 s_3 s_4 s_5 \Gamma^{09}
   \big) \Gamma^{11} \,,
\end{align}
where we have used the definitions of Eq.~(\ref{trig}).
This shows clearly that $1\pm \tilde{\Gamma}$ do not have the form of
projection operators.  One can make them have the desired form by fixing
the transverse angular position, and realize that there is a unique
choice of $\phi^1_0 = \frac{\pi}{2}$ which leads to
$\tilde{\Gamma} = -\Gamma^{04} \Gamma^{11}$.
If we now use the chirality property of $\epsilon_\pm$ in Eq.~(\ref{epch})
and split $\epsilon_\pm$ according to the eigenvalues of $\Gamma^{04}$ as
\begin{align}
\Gamma^{04} \epsilon_{+\pm} = \pm \epsilon_{+\pm} \,, \quad
\Gamma^{04} \epsilon_{-\pm} = \pm \epsilon_{-\pm} \,,
\end{align}
then the worldvolume Killing spinor equation (\ref{kse}) becomes
\begin{align}
0 =& - 2 u^{1/2} \Gamma_{0123} S_0 (\phi)
       \left(
           \epsilon_{+-}
           - x^0 \Gamma_0 \epsilon_{-+}
           - \vec{x}_0 \cdot \vec{\Gamma} \epsilon_{--}
       \right)
\notag \\
 &
    + 2 u^{-1/2} \Gamma_{0123} \Gamma_4 S_0 (\phi) \epsilon_{-+} \,,
\end{align}
with
\begin{align}
S_0(\phi) =  S(\phi)|_{\phi^1=\pi/2} \,.
\end{align}
The solution of this equation is readily found to be
\begin{align}
\epsilon_{-+} = 0 \,, \quad
\epsilon_{+-} = \vec{x}_0 \cdot \vec{\Gamma} \epsilon_{--} \,.
\label{24fix}
\end{align}
Since other components except for those of (\ref{24fix}) are
undetermined, we conclude that the supersymmetry preserved on the
AdS$_2 \times$S$^4$ brane is characterized by
\begin{align}
\epsilon_{++} \,, \quad \epsilon_{--} \,,
\label{24ep}
\end{align}
each of which has eight free components and we have sixteen
supersymmetries (1/2-BPS) in total.

Having identified the worldvolume supersymmetries (\ref{24ep}), it is
straightforward to obtain the supersymmetry transformation rules for the
worldvolume fields according to (\ref{susyt}) and (\ref{betam}).
Firstly, the scalar fields corresponding to the transverse fluctuations
are found to transform as
\begin{align}
\delta \vec{\tilde{x}} &=
    2i u^{-1} \bar{\theta} \vec{\Gamma} \Gamma_{0123}
    (\hat{\eta}_+ - \hat{\eta}_-) + \dots \,,
\notag \\
\delta \tilde{\phi}^1 &=
     2i \bar{\theta} \Gamma_5 \Gamma_{0123}
      (\hat{\eta}_+ - \hat{\eta}_-) + \dots \,.
\end{align}
where
\begin{align}
\hat{\eta}_+ &=
    u^{1/2} S_0(\phi)
    \left(
        \epsilon_{++} - x^0 \Gamma_0 \epsilon_{--}
    \right) \,,
\notag \\
\hat{\eta}_- &=
    u^{-1/2} \Gamma_4 S_0(\phi) \epsilon_{--} \,.
\end{align}
As for the worldvolume gauge field, we obtain
\begin{align}
\delta A_0 &=
    -2i u \bar{\theta} \Gamma_0 \Gamma_{0123}
    (\hat{\eta}_+ - \hat{\eta}_-) + \dots \,,
\notag \\
\delta A_u &=
    -2i u^{-1} \bar{\theta} \Gamma_4 \Gamma_{0123}
     (\hat{\eta}_+ - \hat{\eta}_-) + \dots \,,
\notag \\
\delta A_{\phi^{\alpha}} &=
    -2i e_{\phi^\alpha}^{\hat{a}} \bar{\theta}\Gamma_{\hat{a}}
    \Gamma_{0123} (\hat{\eta}_+ - \hat{\eta}_-) + \dots \,.
\end{align}
where $\alpha = 2,3,4,5$ and $\phi^1 = \frac{\pi}{2}$ should be imposed on
$e_{\phi^\alpha}^{\hat{a}}$. Finally, we get the transformation rule for the
fermionic field as
\begin{align}
\delta \theta &=
   - 2 u \vec{\tilde{x}} \cdot \vec{\Gamma} \Gamma_{01234} \hat{\eta}_-
   + \tilde{\phi}^1 \Gamma_5 \Gamma_{01234}
        (\hat{\eta}_+ + \hat{\eta}_-)
    - \beta^{(5)}_1 (\hat{\eta}_+ + \hat{\eta}_-) + \dots \,,
\end{align}
where
\begin{align}
\beta^{(5)}_1 =& \,
       \bigg(
           \frac{1}{u} \Gamma^0 \partial_0 \tilde{X}^f
           + u \Gamma^4 \partial_u \tilde{X}^f
           + \Gamma^6 \partial_{\phi^2} \tilde{X}^f
           + \frac{1}{s_2} \Gamma^7 \partial_{\phi^3} \tilde{X}^f
\notag \\
 & \,
           + \frac{1}{s_2 s_3}\Gamma^8 \partial_{\phi^4} \tilde{X}^f
           + \frac{1}{s_2 s_3 s_4} \Gamma^9 \partial_{\phi^5} \tilde{X}^f
       \bigg) e_f^{\hat{a}} \Gamma_{046789 \hat{a}}
\notag \\
 &  \,
    - \frac{1}{48 s_2^3 s_3^2 s_4} \epsilon^{i_0\cdots i_3 i_4 i_5}
    \left(
        e^{\hat{a}_0}_{\ell_{i_0}} \cdots e^{\hat{a}_3}_{\ell_{i_3}}
    \right) \bigg|_{\phi^1 = \pi/2}
    \Gamma_{\hat{a}_0 \cdots \hat{a}_3}
    F_{i_4 i_5} \,.
\end{align}

\subsection{D7}

In this subsection, we are led to consider two kinds of D7-brane
configurations.  The common content for them is $\beta^{(7)}_\pm$
appearing in the $\kappa$ symmetry projection $\Gamma^{(7)}$, which is
read off from Eqs.~(\ref{projexp}) and (\ref{beta}) as
\begin{align}
\beta^{(7)}_\pm =
  & \, \frac{\epsilon^{i_1 \cdots i_8}}{\sqrt{-\det (G_{ij}+\mathcal{F}_{ij})}}
    \bigg(
          \pm \frac{1}{8!} \gamma_{i_1 \cdots i_8}
          +   \frac{1}{1440} \gamma_{i_1 \cdots i_6} \mathcal{F}_{i_7 i_8}
          \pm \frac{1}{192}  \gamma_{i_1 \cdots i_4}
                             \mathcal{F}_{i_5 i_6} \mathcal{F}_{i_7 i_8}
\notag \\
  &       +   \frac{1}{96}  \gamma_{i_1 i_2} \mathcal{F}_{i_3 i_4}
                            \mathcal{F}_{i_5 i_6} \mathcal{F}_{i_7 i_8}
          \pm \frac{1}{384} \mathcal{F}_{i_1 i_2} \mathcal{F}_{i_3 i_4}
                            \mathcal{F}_{i_5 i_6} \mathcal{F}_{i_7 i_8}
    \bigg) \,.
\label{b7}
\end{align}

\subsubsection{$(5, 3)$-brane}

The D7-brane configuration (5,3) of Table \ref{tablebps} leads us to take
the static gauge as
\begin{align}
x^{0,1,2,3}(\sigma) = \sigma^{0,1,2,3} \,, \quad
u(\sigma) = \sigma^4 \,, \quad
\phi^{3,4,5} (\sigma) = \sigma^{5,6,7} \,,
\label{53static}
\end{align}
or $(\ell_0, \ell_1, \ell_2, \ell_3, \ell_4, \ell_5, \ell_6, \ell_7) =
(0,1,2,3,4,7,8,9)$ in Eq.~(\ref{static}),
which corresponds to the AdS$_5 \times$S$^3$ brane.  The coordinates transverse
to this configuration are then
\begin{align}
X^f = \{ \phi^1, \phi^2 \} \,.
\end{align}

In the static gauge (\ref{53static}), $\beta^{(7)}_-$ of (\ref{b7}) becomes,
\begin{align}
\beta^{(7)}_0 = \Gamma_{01234789} \,,
\end{align}
according to (\ref{beta0}).  Now
$\left( \Gamma_{0123} \beta^{(7)}_0 \right)^2 = 1$ obviously and thus
$1 \pm \Gamma_{0123} \beta^{(7)}_0$ in the worldvolume Killing spinor
equation (\ref{kse}) with $p=7$ play the role of projection operators.
Having the projection operators, the next step described
in Sec.~\ref{wvsusy} is to send $\Gamma_{0123} \beta^{(7)}_0$ to the
right of $S_0(\phi)$ in (\ref{kse}).  If we denote the resulting
expression as $\tilde{\Gamma}$, we get the relation
$\Gamma_{0123} \beta^{(7)}_0 S_0(\phi)= S_0(\phi) \tilde{\Gamma}$.
With the fact that
$\Gamma_{0123} \beta^{(7)}_0 = -\Gamma_{4789}$, $\tilde{\Gamma}$
is evaluated by repeated use of the identity (\ref{iden}) as follows:
\begin{align}
-\tilde{\Gamma}
&= S_0^{-1} (\phi) \Gamma_{4789} S_0 (\phi)
\notag \\
&= \big(
    \mathring{c}_1 \Gamma_4 + \mathring{s}_1 \mathring{c}_2 \Gamma_5 + \mathring{s}_1 \mathring{s}_2 c_3 \Gamma_6
    + \mathring{s}_1 \mathring{s}_2 s_3 c_4 \Gamma_7
    + \mathring{s}_1 \mathring{s}_2 s_3 s_4 c_5 \Gamma_8
    + \mathring{s}_1 \mathring{s}_2 s_3 s_4 s_5 \Gamma_9 \big)
\notag \\
&\phantom{=\,}
    \times \big(
    c_3 \Gamma_{789} - s_3 c_4 \Gamma_{689} + s_3 s_4 c_5 \Gamma_{679}
    - s_3 s_4 s_5 \Gamma_{678} \big) \,,
\end{align}
where we have used the definitions of Eq.~(\ref{trig}).
This shows clearly that $1\pm \tilde{\Gamma}$ do not have the form of
projection operators.  One can make them have the desired form by fixing
the transverse angular position, and realize that there is a unique
choice of $\phi^{1,2}_0 = \frac{\pi}{2}$ which leads to
$\tilde{\Gamma} = -\Gamma_{6789}$.
If we now split $\epsilon_\pm$ according to the eigenvalues of
$\Gamma_{6789}$ as
\begin{align}
\Gamma_{6789} \epsilon_{+\pm} = \pm \epsilon_{+\pm} \,, \quad
\Gamma_{6789} \epsilon_{-\pm} = \pm \epsilon_{-\pm} \,,
\end{align}
then the worldvolume Killing spinor equation (\ref{kse}) becomes
\begin{align}
0
= & \, - 2 u^{1/2} \Gamma_{0123} S_0(\phi)
           (\epsilon_{++} - x \cdot \Gamma \epsilon_{-+})
\notag \\
  & \, + 2 u^{-1/2} \Gamma_{0123} \Gamma_4 S_0(\phi)
           \epsilon_{-+} \,,
\end{align}
with
\begin{align}
S_0(\phi) =  S(\phi)|_{\phi^1, \phi^2=\pi/2} \,.
\end{align}
The solution of this equation is readily found to be
\begin{align}
\epsilon_{-+} = 0 \,, \quad
\epsilon_{++} = 0 \,.
\label{53fix}
\end{align}
Since other components except for those of (\ref{53fix}) are
undetermined, we conclude that the supersymmetry preserved on the
AdS$_5 \times$S$^3$ brane is characterized by
\begin{align}
\epsilon_{+-} \,, \quad \epsilon_{--} \,,
\label{53ep}
\end{align}
each of which has eight free components and we have sixteen
supersymmetries (1/2-BPS) in total.

Having identified the worldvolume supersymmetries (\ref{53ep}), it is
straightforward to obtain the supersymmetry transformation rules for the
worldvolume fields according to (\ref{susyt}) and (\ref{betam}).
Firstly, the scalar fields corresponding to the transverse fluctuations
are found to transform as
\begin{align}
\delta \tilde{\phi}^{1,2} &=
     2i \bar{\theta} \Gamma^{5,6} \Gamma_{0123}
      (\hat{\eta}_+ - \hat{\eta}_-) + \dots \,,
\end{align}
where
\begin{align}
\hat{\eta}_+ &=
    u^{1/2} S_0(\phi)
       ( \epsilon_{+-} - x \cdot \Gamma \epsilon_{--} ) \,,
\notag \\
\hat{\eta}_- &=
    u^{-1/2} \Gamma_4 S_0(\phi) \epsilon_{--} \,.
\end{align}
As for the worldvolume gauge field, we obtain
\begin{align}
\delta A_{0,1,2,3} &=
    -2i u \bar{\theta} \Gamma_{0,1,2,3} \Gamma_{0123}
       (\hat{\eta}_+ - \hat{\eta}_-) + \dots \,,
\notag \\
\delta A_u &=
    -2i u^{-1} \bar{\theta} \Gamma_4 \Gamma_{0123}
       (\hat{\eta}_+ - \hat{\eta}_-) + \dots \,,
\notag \\
\delta A_{\phi^\alpha} &=
    -2i   e_{\phi^\alpha}^{\hat{a}} \bar{\theta} \Gamma_{\hat{a}}
     \Gamma_{0123} (\hat{\eta}_+ - \hat{\eta}_-) + \dots \,,
\end{align}
where $\alpha = 3,4,5$ and $\phi^{1,2} = \frac{\pi}{2}$ should be imposed on
$e_{\phi^\alpha}^{\hat{a}}$.  Finally, we get the transformation rule for the
fermionic field as
\begin{align}
\delta \theta =
     \tilde{\phi}^\alpha \Gamma_{\alpha+4} \Gamma_{01234}
        (\hat{\eta}_+ + \hat{\eta}_-)
    - \beta^{(7)}_1 (\hat{\eta}_+ + \hat{\eta}_-) + \dots \,,
\end{align}
where $\alpha=1,2$ and
\begin{align}
\beta^{(7)}_1 =& \,
      - \bigg(
           \frac{1}{u} \Gamma^0 \partial_0 \tilde{X}^f
           + \frac{1}{u} \Gamma^1 \partial_1 \tilde{X}^f
           + \frac{1}{u} \Gamma^2 \partial_2 \tilde{X}^f
           + \frac{1}{u} \Gamma^3 \partial_3 \tilde{X}^f
           + u \Gamma^4 \partial_u \tilde{X}^f
\notag \\
 & \,
           + \Gamma^7 \partial_{\phi^3} \tilde{X}^f
           + \frac{1}{s_3}\Gamma^8 \partial_{\phi^4} \tilde{X}^f
           + \frac{1}{s_3 s_4} \Gamma^9 \partial_{\phi^5} \tilde{X}^f
       \bigg) e_f^{\hat{a}} \Gamma_{01234789 \hat{a}}
\notag \\
 &  \,
    + \frac{1}{1440 u^3 s_3^2 s_4} \epsilon^{i_0\cdots i_5 i_6 i_7}
    \left(
        e^{\hat{a}_0}_{\ell_{i_0}} \cdots e^{\hat{a}_5}_{\ell_{i_5}}
    \right) \bigg|_{\phi^1,\phi^2 = \pi/2}
    \Gamma_{\hat{a}_0 \cdots \hat{a}_5}
    F_{i_6 i_7} \,.
\end{align}

\subsubsection{$(3, 5)$-brane}

The D7-brane configuration (3,5) of Table \ref{tablebps} leads us to take
the static gauge as
\begin{align}
x^{0,1}(\sigma) = \sigma^{0,1} \,, \quad
u(\sigma) = \sigma^2 \,, \quad
\phi^{1,\dots,5} (\sigma) = \sigma^{3,\dots,7} \,,
\label{35static}
\end{align}
or $(\ell_0, \ell_1, \ell_2, \ell_3, \ell_4, \ell_5, \ell_6, \ell_7) =
(0,1,4,5,6,7,8,9)$ in Eq.~(\ref{static}),
which corresponds to the AdS$_3 \times$S$^5$ brane.  The coordinates transverse
to this configuration are then
\begin{align}
X^f = \{ x^2,x^3 \} \,.
\end{align}

In the static gauge (\ref{35static}), $\beta^{(7)}_-$ of (\ref{b7}) becomes,
\begin{align}
\beta^{(7)}_0 = \Gamma_{01456789} \,,
\end{align}
according to (\ref{beta0}).  Now
$\left( \Gamma_{0123} \beta^{(7)}_0 \right)^2 = 1$ obviously and thus
$1 \pm \Gamma_{0123} \beta^{(7)}_0$ in the worldvolume Killing spinor
equation (\ref{kse}) with $p=7$ play the role of projection operators.
Having the projection operators, the next step described
in Sec.~\ref{wvsusy} is to send $\Gamma_{0123} \beta^{(7)}_0$ to the
right of $S_0(\phi)$ in (\ref{kse}).  But this process is trivial
because we see that $\Gamma_{0123} \beta^{(7)}_0$ moves freely to the right
of $S_0(\phi)$ by noticing
$\Gamma_{0123} \beta^{(7)}_0 = \Gamma_{23\dots 9} = \Gamma^{01} \Gamma^{11}$
from the definition of $\Gamma^{11}$ given in (\ref{gdef}).
If we now split $\epsilon_\pm$ according to the eigenvalues of
$\Gamma^{01}$ as
\begin{align}
\Gamma^{01} \epsilon_{+\pm} = \pm \epsilon_{+\pm} \,, \quad
\Gamma^{01} \epsilon_{-\pm} = \pm \epsilon_{-\pm} \,,
\end{align}
then the worldvolume Killing spinor equation (\ref{kse}) becomes
\begin{align}
0
= & \, - 2 u^{1/2} \Gamma_{0123} S(\phi)
           \left[ \epsilon_{+-} - (x^0 \Gamma_0 + x^1 \Gamma_1) \epsilon_{-+}
                                - (x_0^2 \Gamma_2 + x_0^3 \Gamma_3)\epsilon_{--}
           \right]
\notag \\
  & \, + 2 u^{-1/2} \Gamma_{0123} \Gamma_4 S(\phi)
            \epsilon_{-+} \,,
\end{align}
The solution of this equation is readily found to be
\begin{align}
\epsilon_{-+} = 0 \,, \quad
\epsilon_{+-} = (x_0^2 \Gamma_2 + x_0^3 \Gamma_3)\epsilon_{--} \,.
\label{35fix}
\end{align}
Since other components except for those of (\ref{35fix}) are
undetermined, we conclude that the supersymmetry preserved on the
AdS$_3 \times$S$^5$ brane is characterized by
\begin{align}
\epsilon_{++} \,, \quad \epsilon_{--} \,,
\label{35ep}
\end{align}
each of which has eight free components and we have sixteen
supersymmetries (1/2-BPS) in total.

Having identified the worldvolume supersymmetries (\ref{35ep}), it is
straightforward to obtain the supersymmetry transformation rules for the
worldvolume fields according to (\ref{susyt}) and (\ref{betam}).
Firstly, the scalar fields corresponding to the transverse fluctuations
are found to transform as
\begin{align}
\delta \tilde{x}^{2,3} &=
     2i u^{-1} \bar{\theta} \Gamma^{2,3} \Gamma_{0123}
        (\hat{\eta}_+ - \hat{\eta}_-) + \dots \,,
\end{align}
where
\begin{align}
\hat{\eta}_+ &=
    u^{1/2} S(\phi)
    \left[
        \epsilon_{++}
        - (x^0 \Gamma_0 + x^1 \Gamma_1 ) \epsilon_{--}
    \right] \,,
\notag \\
\hat{\eta}_- &=
    u^{-1/2} \Gamma_4 S(\phi) \epsilon_{--}
\end{align}
As for the worldvolume gauge field, we obtain
\begin{align}
\delta A_{0,1} &=
    -2i u \bar{\theta} \Gamma_{0,1} \Gamma_{0123}
       (\hat{\eta}_+ - \hat{\eta}_-) + \dots \,,
\notag \\
\delta A_u &=
    -2i u^{-1} \bar{\theta} \Gamma_4 \Gamma_{0123}
       (\hat{\eta}_+ - \hat{\eta}_-) + \dots \,,
\notag \\
\delta A_{\phi^\alpha} &=
    -2i   e_{\phi^\alpha}^{\hat{a}} \bar{\theta} \Gamma_{\hat{a}}
     \Gamma_{0123} (\hat{\eta}_+ - \hat{\eta}_-) + \dots \,,
\end{align}
where $\alpha = 1,\dots,5$.  Finally, we get the transformation rule for the
fermionic field as
\begin{align}
\delta \theta &=
    - 2 u  ( \tilde{x}^2 \Gamma_2 + \tilde{x}^3 \Gamma_3)
       \Gamma_{01234} \hat{\eta}_-
    - \beta^{(7)}_1 (\hat{\eta}_+ + \hat{\eta}_-) + \dots \,,
\end{align}
where
\begin{align}
\beta^{(7)}_1 =& \,
      - \bigg(
           \frac{1}{u} \Gamma^0 \partial_0 \tilde{X}^f
           + \frac{1}{u} \Gamma^1 \partial_1 \tilde{X}^f
           + u \Gamma^4 \partial_u \tilde{X}^f
           + \Gamma^5 \partial_{\phi^1} \tilde{X}^f
           + \frac{1}{s_1} \Gamma^6 \partial_{\phi^2} \tilde{X}^f
\notag \\
 & \,
           + \frac{1}{s_1 s_2}\Gamma^7 \partial_{\phi^3} \tilde{X}^f
           + \frac{1}{s_1 s_2 s_3} \Gamma^8 \partial_{\phi^4} \tilde{X}^f
           + \frac{1}{s_1 s_2 s_3 s_4} \Gamma^9 \partial_{\phi^5} \tilde{X}^f
       \bigg) e_f^{\hat{a}} \Gamma_{01456789 \hat{a}}
\notag \\
 &  \,
    + \frac{1}{1440 u s_1^4 s_2^3 s_3^2 s_4} \epsilon^{i_0\cdots i_5 i_6 i_7}
        e^{\hat{a}_0}_{\ell_{i_0}} \cdots e^{\hat{a}_5}_{\ell_{i_5}}
    \Gamma_{\hat{a}_0 \cdots \hat{a}_5}
    F_{i_6 i_7} \,.
\end{align}

\subsection{Invariance of quadratic action}

The transformation rules obtained in the previous subsections are
explicit to the leading linear order in the worldvolume fluctuating
fields.  This means that they can be used to confirm the invariance
of the quadratic action coming from the expansion of the gauge fixed
D-brane action in terms of the worldvolume fields.  In this last
subsection, we would like to verify the transformation rules by
showing the invariance of the quadratic action.  However, we will
not consider all the six kinds of AdS branes but take one representative,
since the transformation rules have the same pattern.

We consider the (3,1) configuration of D3-brane of Sec.~\ref{31config},
the AdS$_3 \times$S$^1$ brane, as the representative.  The D3-brane
action is given in Eq.~(\ref{d3action}) for $p=3$ and the closed
five-form $H_5$ in the WZ term \cite{Metsaev:1998hf} is
\begin{eqnarray}
H_5
&=& - \frac{i}{6} L^{\hat{a}} \wedge L^{\hat{b}} \wedge L^{\hat{c}}
      \wedge \bar{L}^I \wedge
      \Gamma^{\hat{a}\hat{b}\hat{c}} \tau_2^{IJ} L^J
    - i \mathcal{F} \wedge L^{\hat{a}} \wedge
        \bar{L}^I \wedge \Gamma^{\hat{a}} \tau_1^{IJ} L^J
\notag \\
& & + \frac{1}{30}
    \left(  \epsilon^{ a_1 \dots a_5} L^{a_1} \wedge \dots \wedge L^{a_5}
       +\epsilon^{a_1' \dots a_5'} L^{a_1'} \wedge \dots \wedge L^{a_5'}
    \right) \,.
\end{eqnarray}
Then, in the static gauge of (\ref{31static}) and the covariant $\kappa$
symmetry fixing condition (\ref{covgauge}), the expansion of the
action in terms of the fluctuating fields leads us to have the bosonic
($S^{(2)}_B$) and fermionic ($S^{(2)}_F$) parts of the quadratic action as
\begin{align}
S^{(2)}_B =& \int d^4 \sigma \sqrt{-g}
    \bigg[
        - \frac{1}{2} u^2 g^{ij}
             \partial_i \tilde{x}^m \partial_j \tilde{x}^m
         - 2 u^2 (\tilde{x}^3 \partial_{\phi^5} \tilde{x}^2
                   - \tilde{x}^2 \partial_{\phi^5} \tilde{x}^3 )
\notag \\
     &   - \frac{1}{2} g^{ij}
            \partial_i \tilde{\phi}^\alpha \partial_j \tilde{\phi}^\alpha
        + \frac{1}{2} ( \tilde{\phi}^\alpha )^2
        - \frac{1}{4} g^{ij}g^{kl} F_{ik}F_{jl}
    \bigg] \,,
\notag \\
S^{(2)}_F = & \,\, i \int d^4 \sigma \sqrt{-g}
    \left(
        e^i_{\bar{a}} \bar{\theta} \Gamma^{\bar{a}} \nabla_i \theta
      + \bar{\theta} \Gamma_{239} \theta
    \right) \,,
\label{2action}
\end{align}
where $m=2,3$, $\alpha=1,2,3,4$, $\bar{a}=0,1,4,9$, and the terms linear in
the derivative $\partial_{\phi^5}$ and the fermionic mass term originate from
the WZ term.  The metric $g_{ij}$ is the induced one on the worldvolume
given by
\begin{align}
g_{ij} d\sigma^i d \sigma^j =
    - u^2 (d x^0)^2 + u^2 (d x^1)^2 + \frac{du^2}{u^2} + (d \phi^5)^2 \,.
\end{align}
This induced metric is also expressed as
$e_i^{\bar{a}}e_j^{\bar{b}} \eta_{\bar{a}\bar{b}}$, where $e_i^{\bar{a}}$
is defined by the worldvolume field independent part of the pullback of
zehnbein (\ref{pbzb}), that is,
$e_i^{\bar{a}} \equiv \partial_i X^\mu e_{\mu}^{\bar{a}}|_{\tilde{X}^f=0}$.
In the covariant derivative for the spinor given by
$\nabla_i = \partial_i + \frac{1}{4} \omega_i^{\bar{a}\bar{b}}
\Gamma_{\bar{a}\bar{b}}$, the spin connection is determined from
$e_i^{\bar{a}}$ by using the Cartan structure equation or can be defined,
like the definition of $e_i^{\bar{a}}$, as
$\omega_i^{\bar{a}\bar{b}} \equiv \partial_i X^\mu
\omega_\mu^{\bar{a}\bar{b}}|_{\tilde{X}^f=0}$ from the spacetime
spin connection $\omega_\mu^{\bar{a}\bar{b}}$.  In the present case,
the nonvanishing components are $\omega^{04}=u dx^0$ and
$\omega^{14} = u dx^1$.

We note that the Lagrangian density for $\tilde{x}^m$ in the quadratic action
(\ref{2action}) is not of the canonical form because of the overall $u^2$
factor.  In order to make it canonical, we take the rescaling of $\tilde{x}^m$
as
\begin{align}
\tilde{x}^m \,\, \longrightarrow \,\, \frac{\tilde{x}^m}{u} \,,
\end{align}
under which the Lagrangian density for the field $\tilde{x}^m$ becomes
\begin{align}
 - \frac{1}{2} g^{ij}
             \partial_i \tilde{x}^m \partial_j \tilde{x}^m
- \frac{3}{2} (\tilde{x}^m)^2
- 2 (\tilde{x}^3 \partial_{\phi^5} \tilde{x}^2
                   - \tilde{x}^2 \partial_{\phi^5} \tilde{x}^3 ) \,.
\label{scaledaction}
\end{align}
There are also associated changes in the transformation rules of
Eqs.~(\ref{xtransf}) and (\ref{stransf}) as
\begin{align}
\delta \tilde{x}^m &=
    2i \bar{\theta} \Gamma^m \Gamma_{0123}
    (\hat{\eta}_+ - \hat{\eta}_-) + \dots \,,
\notag \\
\delta \theta &=
    - 2 \tilde{x}^m \Gamma_m \Gamma_{01234} \hat{\eta}_-
      + \tilde{\phi}^\alpha \Gamma_{\alpha+4} \Gamma_{01234}
        (\hat{\eta}_+ + \hat{\eta}_-)
    - \beta^{(3)}_1 (\hat{\eta}_+ + \hat{\eta}_-) + \dots \,,
\label{scaledt}
\end{align}
where $\tilde{x}^m$ dependent part in $\beta^{(3)}_1$ of (\ref{beta31}) becomes
\begin{align}
 - e^i_{\bar{a}} \Gamma^{\bar{a}} \partial_i \tilde{x}^m
   \Gamma_{0149} \Gamma_m
   + \tilde{x}^m \Gamma_{019} \Gamma_m \,.
\label{scaledb}
\end{align}

Bearing in mind the above rescaled expressions (\ref{scaledaction}),
(\ref{scaledt}) and (\ref{scaledb}), we are now ready to consider the
transformation of the quadratic action (\ref{2action}) by applying
the transformation rules (\ref{xtransf}), (\ref{gtransf}) and (\ref{stransf})
with (\ref{d3eta}) and (\ref{beta31}).  The calculation itself is less trivial
but straightforward, and the final result is
\begin{align}
\delta S^{(2)}_B + \delta S^{(2)}_F = 0 \,.
\end{align}
This invariance of the quadratic action clearly shows that the
transformation rules realize the supersymmetry of the worldvolume theory.

\section{Non-AdS branes}
\label{nonadsbrane}

We turn to the non-AdS branes, in which the AdS radial direction $u$ is
transverse to the D-brane configuration and acts as a worldvolume field.
From Table \ref{tablebps}, it turns out that there are six types of Lorentzian
non-AdS D-brane configurations, which are supposed to be supersymmetric.
For these configurations, the brane embedding coordinates $X^{\ell_i}$ chosen
by the static gauge condition (\ref{static}) and the associated
$\beta_0^{(p)}$ of (\ref{beta0}) are listed in Table \ref{table-nonads}.

\begin{table}
\begin{center}
\begin{tabular}{c|ccc}
\hline
D$p$ & ($n$,$n'$) & $X^{\ell_i}$ & $\beta^{(p)}_0$  \\ \hline\hline
D1   & (2,0)      & $x^{0,1}$    & $-\Gamma_{01}$
\\ \hline
D3   & \begin{tabular}{c} (3,1) \\ (1,3) \end{tabular} &
       \begin{tabular}{c} $x^{0,1,2}$, $\phi^5$ \\
                          $x^0$, $\phi^{3,4,5}$
       \end{tabular} &
       \begin{tabular}{c} $\Gamma_{0129}$ \\ $\Gamma_{0789}$
       \end{tabular}
\\ \hline
D5   & \begin{tabular}{c} (4,2) \\ (2,4) \end{tabular} &
       \begin{tabular}{c} $x^{0,1,2,3}$, $\phi^{4,5}$ \\
                          $x^{0,1}$, $\phi^{2,3,4,5}$
       \end{tabular} &
       \begin{tabular}{c} $-\Gamma_{012389}$ \\ $-\Gamma_{016789}$
       \end{tabular}
\\ \hline
D7   & (3,5)     & $x^{0,1,2}$, $\phi^{1,\dots,5}$ & $\Gamma_{01256789}$
\\ \hline
\end{tabular}
\caption{\label{table-nonads} Configurations of Non-AdS branes}
\end{center}
\end{table}

As described in Sec.~\ref{wvsusy}, a given D-brane configuration is
supersymmetric if the eigenvalues of $\Gamma_{0123} \beta^{(p)}_0$ are
$\pm 1$, since $1 \pm \Gamma_{0123} \beta^{(p)}_0$ appearing in the
worldvolume Killing spinor equation (\ref{kse}) are then projection
operators and hence used to pick out preserved supersymmetries from
$\epsilon_+$ and $\epsilon_-$.  For all the expressions of
$\beta^{(p)}_0$ listed in Table~\ref{table-nonads}, however,
$\left( \Gamma_{0123} \beta^{(p)}_0 \right)^2 = -1$ which means that
the eigenvalues of $\Gamma_{0123} \beta^{(p)}_0$ are $\pm i$.  Therefore,
we conclude that the static non-AdS D-branes are not supersymmetric.

The radial AdS coordinate in the worldvolume Killing spinor equation
(\ref{kse}) should be understood as $u_0$, the AdS radial position of
D-brane.  One may argue that if a non-AdS D-brane is placed at the origin
of $u$, that is $u_0=0$, then the solution of the Killing spinor equation
is $\epsilon_- = 0$ while leaving $\epsilon_+$ free and thus the non-AdS
brane at such special position is supersymmetric.  The problem is that the
induced worldvolume metric becomes singular at $u_0=0$.  To avoid the
singularity, one might consider a kind of regularization by introducing a
non-vanishing infinitesimal value of $u_0$.  However, he or she encounters
again the fact that $1 \pm \Gamma_{0123} \beta^{(p)}_0$ are not projection
operators.  Thus the conclusion about the supersymmetry of non-AdS
branes does not change.

It should be noted that the conclusion in this section is only for the
static configurations.  If the non-AdS brane takes a certain constant motion
along a transverse direction or has non-vanishing fluxes on its worldvolume,
the situation may change completely.  A typical example is the giant graviton
\cite{McGreevy:2000cw}, one type of which is a (1,3) configuration of D3 brane
and takes a constant motion along a transverse angular direction.  It is known
to be 1/2-BPS for some particular angular speed.

\section{Discussion}
\label{concl-disc}

Starting from the data in Table \ref{tablebps}, the classification of 1/2-BPS
D-branes on the AdS$_5 \times$S$^5$ background obtained by the covariant open
string description, we have considered all possible static D-brane
configurations without worldvolume fluxes.  We have identified which part of
the target spacetime supersymmetry is preserved on the D-brane worldvolume by
solving the worldvolume Killing spinor equation and showed that only the
AdS type D-branes where the AdS radial direction is a worldvolume coordinate
are 1/2-BPS.  As for the supersymmetric configurations, we have obtained the
associated worldvolume supersymmetry transformation rules for the worldvolume
fields.

One interesting point in the study of supersymmetric configuration is that the
transverse angular position has been determined without resort to the
equations of motion.  The position is fixed only by requiring the
worldvolume supersymmetry and sometimes the non-degeneracy of induced
worldvolume metric.  For example, let us consider the AdS$_4 \times$S$^2$
embedding of D5-brane of Sec.~\ref{42config}, which was also explored in
Ref.~\cite{Skenderis:2002vf} related to the holographic description
of the defect conformal field theory \cite{DeWolfe:2001pq}.  As shown in
\cite{Skenderis:2002vf}, there are two solutions of the equations of motion
for the transverse angular position and it turns out that only one of them
leads to the supersymmetric configuration which is the same as the angular
position determined in this paper as it should be.  In fact, this kind of
situation seems to be natural.  Usually, the solution of the Killing spinor
equation satisfies also the equations of motion.  Thus the fact that the
transverse angular position is determined in the process of solving the
Killing spinor equation may not be surprising.

As shown in Sec.~\ref{nonadsbrane}, all the static non-AdS branes without any
worldvolume fluxes are not supersymmetric.  We emphasize again that this
may change completely when there are motions in transverse directions or
the worldvolume fluxes are turned on, since we already know at least one
definite example, the giant graviton.  Beyond the static case, there will
be lots of possibilities.  Having said that, we expect that there will be
a suitable classification facilitating the study of them similar to
the static D-brane configurations classified as the AdS and non-AdS types.

\section*{Acknowledgments}

This work was supported by Basic Science Research Program through the National Research Foundation of Korea (NRF) funded by the Ministry of Science, ICT
and Future Planning with Grant No.~NRF-2015R1A2A2A01007058 (JP) and
NRF-2015R1A2A2A01004532 (HS).

\appendix
\section{Notation and convention}
\label{app}

The notation for the supercoordinate we use is
\begin{align}
Z^M = (X^\mu, \Theta^I) \,,
\end{align}
where the spinor index for the fermionic coordinate $\Theta$ has been
suppressed, $\mu$ is the ten-dimensional curved space-time vector index,
and $I$ ($=1,2$) is introduced to distinguish the two
spinors with the  same chirality (The chirality is taken to be positive).
As for the Lorentz frame or the tangent space, the vector index is
denoted by
\begin{align}
\hat{a} = (a, a') \,, \quad a=0,1,2,3,4 \,, \quad a'=5,\dots,9 \,,
\end{align}
where $a$ ($a'$) corresponds to the tangent space of AdS$_5$ ($S^5$), and
the metric $\eta_{\hat{a}\hat{b}}$ follows the most plus sign convention as
$\eta_{\hat{a}\hat{b}} = \text{diag} (-, +, +, \dots, +)$.

The matrices acting on the spinors indexed with $I,J,\dots$
are denoted by
\begin{align}
\tau_1 = \begin{pmatrix} 0 & 1 \\ 1 & 0 \end{pmatrix} \,, \quad
\tau_2 = \begin{pmatrix} 0 & 1 \\ -1 & 0 \end{pmatrix} \,, \quad
\tau_3 = \begin{pmatrix} 1 & 0 \\ 0 & -1 \end{pmatrix} \,.
\label{taumat}
\end{align}
The explicit expression for the vector (spinor) superfield or the
Maurer-Cartan one-form superfield
$L^{\hat{a}} = dZ^M L^{\hat{a}}_M$ ($L^I = dZ^M L_M^I$) is given by
\cite{Metsaev:1998it,Kallosh:1998zx}
\begin{align}
L^{\hat{a}} &= e^{\hat{a}} - 2 i \sum_{n=0}^{15} \frac{1}{(2n+2)!}
                 \bar{\Theta}^I \Gamma^{\hat{a}}
                 ({\mathcal M}^{2n})^{IJ} D\Theta^J  \,,
\notag \\
L^I &= \sum^{16}_{n=0} \frac{1}{(2n+1)!} (\mathcal{M}^{2n})^{IJ}
      D\Theta^J
\,,
\label{superfld}
\end{align}
where $\Gamma^{\hat{a}}$ is the $32 \times 32$ Dirac gamma matrix, and
${\mathcal M}^2$ and the spinor covariant derivative $D\Theta^I$
are, in the 32 component notation,
\begin{gather}
({\mathcal M}^2)^{IJ}
 = - \epsilon^{IK} \Gamma_* \Gamma^{\hat{a}} \Theta^K
     \bar{\Theta}^J \Gamma_{\hat{a}}
   + \frac{1}{2} {\tau_2}^{KJ}
       ( \Gamma^{ab} \Theta^I \bar{\Theta}^K \Gamma_{ab} \Gamma_*
        -\Gamma^{a'b'} \Theta^I \bar{\Theta}^K \Gamma_{a'b'} \Gamma_*') \,,
\notag \\
D\Theta^I
= \left( d + \frac{1}{4} \omega^{\hat{a}\hat{b}}\Gamma_{\hat{a}\hat{b}} \right) \Theta^I
   - \frac{i}{2} {\tau_2}^{IJ} e^{\hat{a}} \Gamma_* \Gamma_{\hat{a}} \Theta^J \,.
\label{m2dt}
\end{gather}
Some definitions of gamma matrix products and their properties are
as follows:
\begin{gather}
\Gamma_* \equiv i \Gamma^{01234} \,, \quad
\Gamma_*' \equiv -i \Gamma^{56789} \,, \quad
\Gamma_*^2 = 1 \,, \quad \Gamma_*'^2=-1 \,,
\notag \\
\Gamma^{11} = \Gamma^{01\dots 9} = \Gamma_* \Gamma_*' \,, \quad
(\Gamma^{11})^2 = 1 \,.
\label{gdef}
\end{gather}


\end{document}